\documentclass[hyperref]{elsart4-1}

\usepackage{graphicx}
\usepackage{epsfig}
\usepackage{amssymb}

\usepackage[english,francais]{babel}
\usepackage[utf8]{inputenc}
\usepackage{unites2e}
\usepackage[numbers]{natbib}
\usepackage{color}
\usepackage{warn}
\usepackage{wrapfig}

\newtheorem{e-proposition}[theorem]{Proposition}

\newtheorem{e-definition}[theorem]{Definition\rm}

\renewcommand{\theequation}{\arabic{equation}}
\def\grays{$\gamma$~rays}
\def\gray{$\gamma$~ray}

\def\og{\leavevmode\raise.3ex\hbox{$\scriptscriptstyle\langle\!\langle$~}}
\def\fg{\leavevmode\raise.3ex\hbox{~$\!\scriptscriptstyle\,\rangle\!\rangle$}}

\begin{document}

\centerline{Astrophysics}
\begin{frontmatter}



\selectlanguage{english}
\title{Ground-based detectors in very-high-energy gamma-ray astronomy}


\selectlanguage{english}
\author[LLR]{Mathieu de Naurois}
\ead{denauroi@in2p3.fr}
\author[Daniel]{Daniel Mazin}
\ead{mazin@mpp.mpg.de}

\address[LLR]{Laboratoire Leprince-Ringuet, Ecole polytechnique \\
91128 PALAISEAU Cedex, France }
\address[Daniel]{Institute for Cosmic Ray Research, University of Tokyo \\
277-8582 Chiba, Japan }

\medskip

\begin{abstract}

\vskip 0.5\baselineskip
Following the discovery of the cosmic rays by Victor Hess in 1912, more than 70 years and
numerous technological developments were needed before an unambiguous detection of the first very-high-energy gamma-ray source
in 1989 was made. Since this discovery the field on very-high-energy gamma-ray astronomy experienced a true revolution:
A second, then a third generation of instruments were built, observing the atmospheric cascades from the ground,
either through the atmospheric Cherenkov light they comprise, or via the direct detection of the charged particles
they carry. Present arrays, 100 times more sensitive than the pioneering experiments, have detected a large
number of astrophysical sources of various types, thus opening a new window on the non-thermal Universe.
New, even more sensitive instruments are currently being built; these will allow us to explore further this
fascinating domain. In this article we describe the detection techniques, the history of the field and the
prospects for the future of ground-based very-high-energy gamma-ray astronomy.
\vskip 0.5\baselineskip

\selectlanguage{francais}
\noindent{\bf R\'esum\'e: Détecteurs au sol en Astronomie Gamma de Très Haute Energie }

\vskip 0.5\baselineskip

\vskip 0.5\baselineskip
Depuis la découverte des rayons cosmiques en 1912 par Victor Hess, il aura fallu près de 70 ans et de
nombreux développements, pour aboutir à la première détection d'une source gamma de très haute énergie
en 1989. Depuis cette découverte, le domaine de l'astronomie gamma de très haute énergie a vécu
une véritable révolution: des détecteurs de deuxième, puis de troisième génération ont vu le jour, observant
les cascades atmosphériques depuis le sol, soit à travers l'émission Tcherenkov atmosphérique qui les
accompagne, soit en détectant directement les particules chargées qui les composent. Les réseaux
récents, environ 100 fois plus sensibles que les plus anciens, ont détecté de très nombreuses sources astrophysiques 
de types variés et ont ainsi ouvert une nouvelle fenêtre sur l'Univers non thermique. 
De nouveaux réseaux de télescopes encore plus sensibles, en cours de construction, vont nous permettre
de pousser encore plus loin l'exploration de ce domaine fascinant.
Dans cet article, nous décrivons les techniques de détection, dressons un panorama historique du domaine et présentons
les perspectives pour le futur de l'astronomie gamma de très haute énergie au sol.
\vskip 0.5\baselineskip
\noindent{\small{\it Keywords~:} Gamma-rays; Cherenkov Detectors; Imaging Atmospheric Cherenkov Telescopes \vskip 0.5\baselineskip
\noindent{\small{\it Mots-cl\'es~:} Rayons Gammas~; Détecteurs Tcherenkov; Télescopes à effet Cherenkov Atmosphériques}}

\end{abstract}
\end{frontmatter}


\selectlanguage{english}

\section{Introduction - Atmospheric Showers}
\label{sec:intro}

The field of very-high energy (VHE) gamma-ray astronomy has been 
intimately linked to the physics of cosmic rays (CRs) since the discovery of the latter in 1912.
Indeed, it was rapidly noticed that processes giving rise to non-thermal, very-high-energy particles would also lead, via
the interaction of those particles with the interstellar medium (matter and radiation),
to the production of very high energy photons~\cite{CRAS_INTRO}.

Although the first attempts to detect the Cherenkov light from the charged particles traveling in the
atmosphere dates back from 1953~\cite{1953Natur.171..349G}, after a suggestion from Blackett~\cite{1948esns.conf...34B}, 
it took several decades before the emergence of ground-based very-high-energy gamma-ray astronomy. 
Before even trying to distinguish the gamma rays from the charged cosmic rays, 
the main challenges to overcome at that time were to actually detect a Cherenkov signal itself.
The difficulties in the detection were caused by the very short duration of the flashes, the small intensity of the signal and the very
large background from the night sky (light from stars and scattered light, that required the use
of sensitive detectors and fast electronics.

In 1989, the first source of VHE gamma rays was 
discovered by the Whipple collaboration~\cite{whipple-crab}.
This seminal detection opened a new window in gamma-ray astronomy and started a very productive research field
in an energy domain which is essentially accessible only to ground-based instruments.

In this paper, we discuss various concepts of detecting gamma rays
from the Earth's surface as well as major ground-based experiments of the past, 
present and near future that largely shaped and continue to shape
the booming field of gamma-ray astronomy.

\subsection{Development of atmospheric Showers}

Ever since 1912, when Victor Hess announced the first evidence 
that ionizing radiation constantly impinges on the Earth's atmosphere,
scientists are continuing to develop efficient techniques 
to detect and study this radiation.

When a high-energy particle (\gray\ or charged nucleus) enters the atmosphere, it can interact
with the atmospheric nuclei through various processes, leading to the development of a 
so-called ``{\it extended air shower} (EAS)'' of particles.
{\it Electromagnetic showers}, initiated by high energy photons or electrons, are governed by mainly two
elementary processes:

\begin{itemize}
\item Production of pairs of $e^\pm$ by the conversion of high energy photons in the Coulomb field of the nuclei,
\item Bremsstrahlung emission of $e^\pm$ in the same Coulomb field, leading the production of further high energy photons.
\end{itemize}

The energy of the impinging particle is then redistributed over many particles as the shower develops in the atmosphere.
Pair production and bremsstrahlung emission have the same characteristic length, the ``{\it electromagnetic radiation length}'', defined for a material
of mass and atomic numbers $A$ and $Z$ as:

\begin{equation}
X_{0}= \left[ {4\alpha
r_{e}^{2}\frac{N_{A}}{A}Z^{2}\ln\left(183\, Z^{-1/3}\right)}\right]^{-1}\;\left[\U{g}\UU{cm}{-2}\right]
\end{equation}

\noindent where $\alpha = 1/137$ is the fine structure constant, $r_e$ the classical electron radius and $N_A$ the Avogadro number.
This quantity, expressed as a {\it density-integrated thickness} $X = \int\!\!\rho \, \mathrm{d}z$, represents roughly the amount of traversed 
matter after which an electron loses a significant fraction of its energy by bremsstrahlung. The bremsstrahlung emission leads to an average energy loss
as a function of thickness $X$:

\begin{equation}
E(X)=E_{0}\exp \left(-\frac{X}{X_{0}}(1+b)\right)
\end{equation}

\noindent where $\displaystyle b=1 / \left(18\ \ln (183/Z^{1/3}) \right)=0,0122$ in air, so that, on average, each electron looses half of its energy after a depth $R=X_{0}\ln 2$ .
Similarly, the integrated pair-creation probability is given by:

\begin{equation}
\mu (X)=1- \exp \left(-\frac{X}{X_{0}}\left(\frac{7}{9}-\frac{b}{3}\right)\right)
\end{equation}

In the atmosphere (dry air), the radiation length corresponds to $36.7\U{g}\UU{cm}{-2}$. The atmosphere is
therefore a thick calorimeter of $\sim 27$ radiation lengths compared to about $\sim 10 X_0$ for gamma-ray satellites
and $\sim 25 X_0$ for particle physics calorimeter such as those of ATLAS or CMS at the CERN Large Hadron Collider.

\begin{figure}[t!]
\begin{center}
\includegraphics[width=0.7\textwidth]{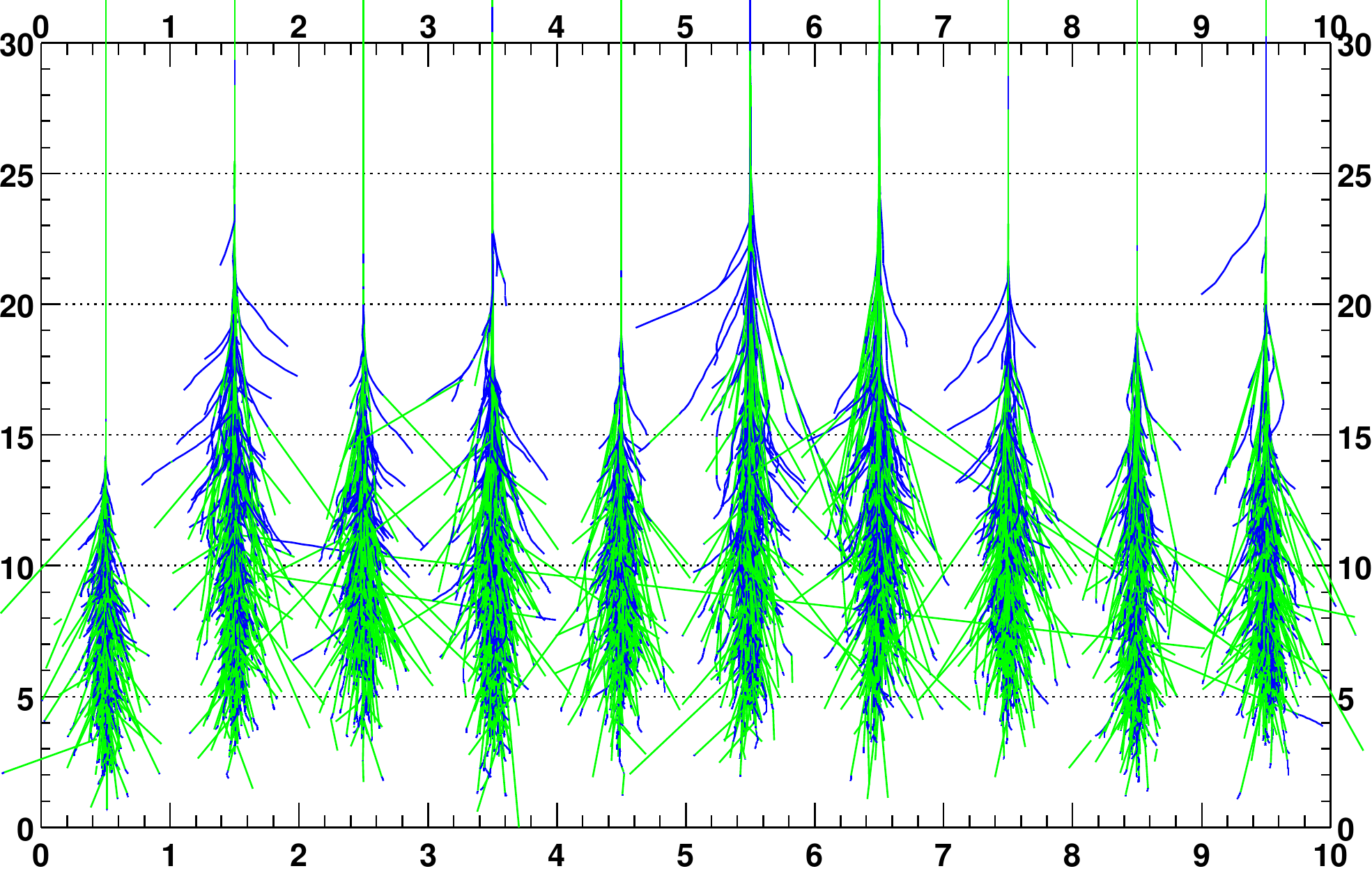} \\
\includegraphics[width=0.7\textwidth]{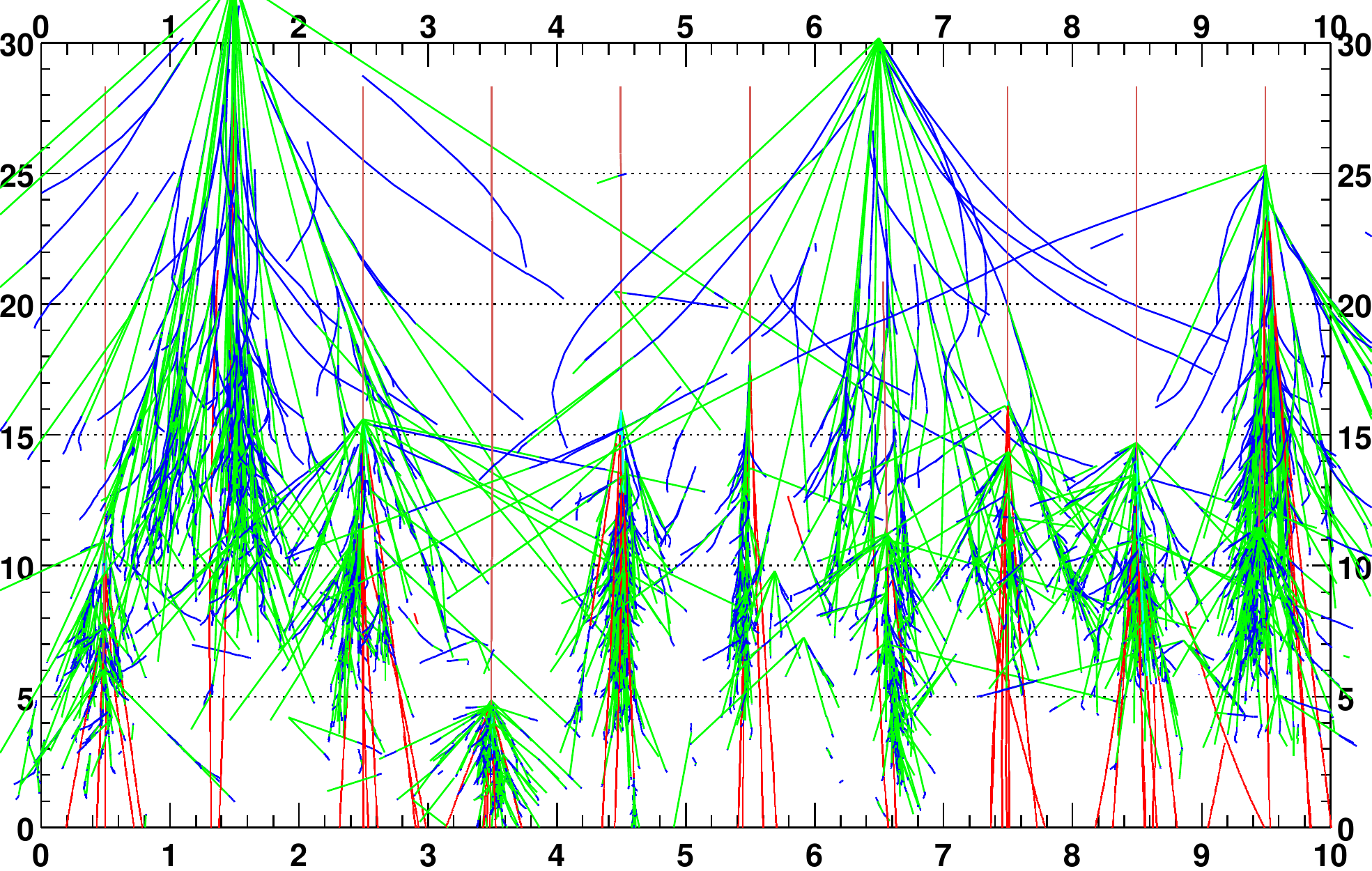} 
\end{center}
\caption{\label{fig:Shower2}Illustration of the intrinsic variability of shower development. {\bf Top:} Simulation
of 10 showers, each initiated by a \gray\ of $300\U{GeV}$. {\bf Bottom:} Simulation of 10 showers initiated by a proton of the same energy.
Due to larger transverse momentum transfers, hadronic showers show larger fluctuations.
From~\cite{TheseMathieu}.}
\end{figure}

It has to be noted that, due to its varying density, the atmosphere is a strongly inhomogeneous calorimeter. At sea level,
for an atmospheric density of $\sim 1.2 \U{kg}\UU{m}{-3}$, one radiation length corresponds to $\sim 300\U{m}$. At $10\U{km}$ 
altitude (roughly the altitude of  the maximum of development of the showers), the radiation length corresponds to a 3 times larger distance ($\sim 1\U{km}$).
This has an important consequence: as a shower penetrates deeper in the atmosphere, its development accelerates due
to the larger amount of target matter. In a homogeneous calorimeter, the depth of the shower maximum depends logarithmically 
on the energy of the primary particle. In the atmosphere, the evolution of the altitude of the shower maximum is even slower. 
This can easily be shown in the framework of the simplified model of an isothermal atmosphere in hydrostatic equilibrium. Then, one finds:

\begin{equation}
z_\mathrm{max} = z_0 \log \frac{\rho_0 z_0}{X_0} - z_0 \log \left( \log \left(\frac{E_0}{E_c}\right) \right) \approx 9\U{km} - 8.4 \U{km} \times \log \left( \log \left(\frac{E_0}{1\U{TeV}}\right) \right)
\end{equation}

where $z_0 = RT/gM \approx 8.4 \U{km}$, $T$ being the absolute temperature, $R$ the perfect gas constant, $M$ the equivalent molar mass for air and $g$ the gravity acceleration. 
The conclusion remains valid in a more realistic model of the atmosphere.

Additional processes play a significant role in the shower development, mainly at low energy:

\begin{itemize}
\item Multiple scattering of charged particles, leading to shower broadening;
\item Energy losses of  $e^\pm$ by ionization and atomic excitation, leading to rapid extinction of the shower when the energy
of the charged particles in the shower pass below the so-called ``{\it critical energy}''\footnote{The critical energy is the energy where the
energy losses by ionization are equal to that by bremsstrahlung. Below this energy, the ionization losses rise as $1/E$ as the particle decelerate,
leading to a very rapid extinction (``{\it Bragg peak}'').} ($E_c = 83\U{MeV}$ in the air);
\item Electron scattering and positron annihilation which leads to an excess of $\sim 10\%$ of electrons compared to positrons (``charge excess''),
which in turn can produce a significant radio emission signal (``{\it Askaryan effect}''). 
\item The Earth's magnetic field which broadens the shower in the East-West direction.
\end{itemize}

At high energy, photo-production or electro-production of hadrons can occasionally give rise to a hadronic component in electromagnetic showers.
However the corresponding cross-sections are typically a factor or $10^{-3}$ smaller than that of pair creation.

Hadronic showers are more complicated to describe, and depend on several different characteristic lengths (nuclear interaction length, 
 decay lengths for unstable particles, radiation length) so no universal scaling is applicable. They comprise several components:

\begin{itemize}
\item Hadronic components: nuclear fragments resulting from collision with atmospheric nuclei, isolated nucleons, $\pi$ and $K$ mesons, etc.,
\item Electromagnetic component resulting in particular from the decay of neutral pions into \grays,
\item High energy muons resulting from the decay of charged mesons ($\pi^\pm$ and $K^\pm$ mainly),
\item Atmospheric neutrinos resulting from the decay of mesons and muons ($\pi^\pm$, $K^\pm$ and $\mu^\pm$).
\end{itemize}

The electromagnetic and hadronic showers are illustrated in Fig.~\ref{fig:Shower2}. Hadronic showers
are more irregular, often comprising several electromagnetic sub-showers.

\subsection{Semi-analytic model of electromagnetic showers}

In the 1950's, 
Greisen~\cite{Greisen1956} proposed a semi-empirical model of the electromagnetic shower development that, in particular,
takes into account ionization losses which where neglected in the previous models.

This models introduces a {\it shower age} parameter, which depends on the primary energy $E_0$, the critical energy $E_{c}$
and the reduced depth $t = X / X_0$:

\begin{equation}
s=\frac{3t}{t+2y}\hbox{ with } y=\ln \left(\frac{E_{0}}{E_{c}}\right)
\end{equation}

The age is $s=0$ at the start of the shower,  $s=1$ at the depth of shower maximum 
$t_{\mathit{max}}=y=\ln (E_{0}/E_{c})$ and $s>1$ in the following extinction phase.
The semi-empirical Greisen formula gives the average number of electrons at
depth $t$ and at the depth of shower maximum $t_{\mathit{max}}$ respectively:

\begin{equation}
 \overline{{N_{e}}}(t)=\frac{0,31}{\sqrt{y}}\exp\left[t\left(1-\frac{3}{2}\ln s\right)\right]
~~ \longrightarrow ~~ \overline{{N_{e}}}(t_{\mathit{max}})=\frac{0,31}{\sqrt{y}}\left(\frac{E_{0}}{E_{c}}\right)
\end{equation}

\begin{figure}[t!]
\begin{center}
\includegraphics[width=\textwidth]{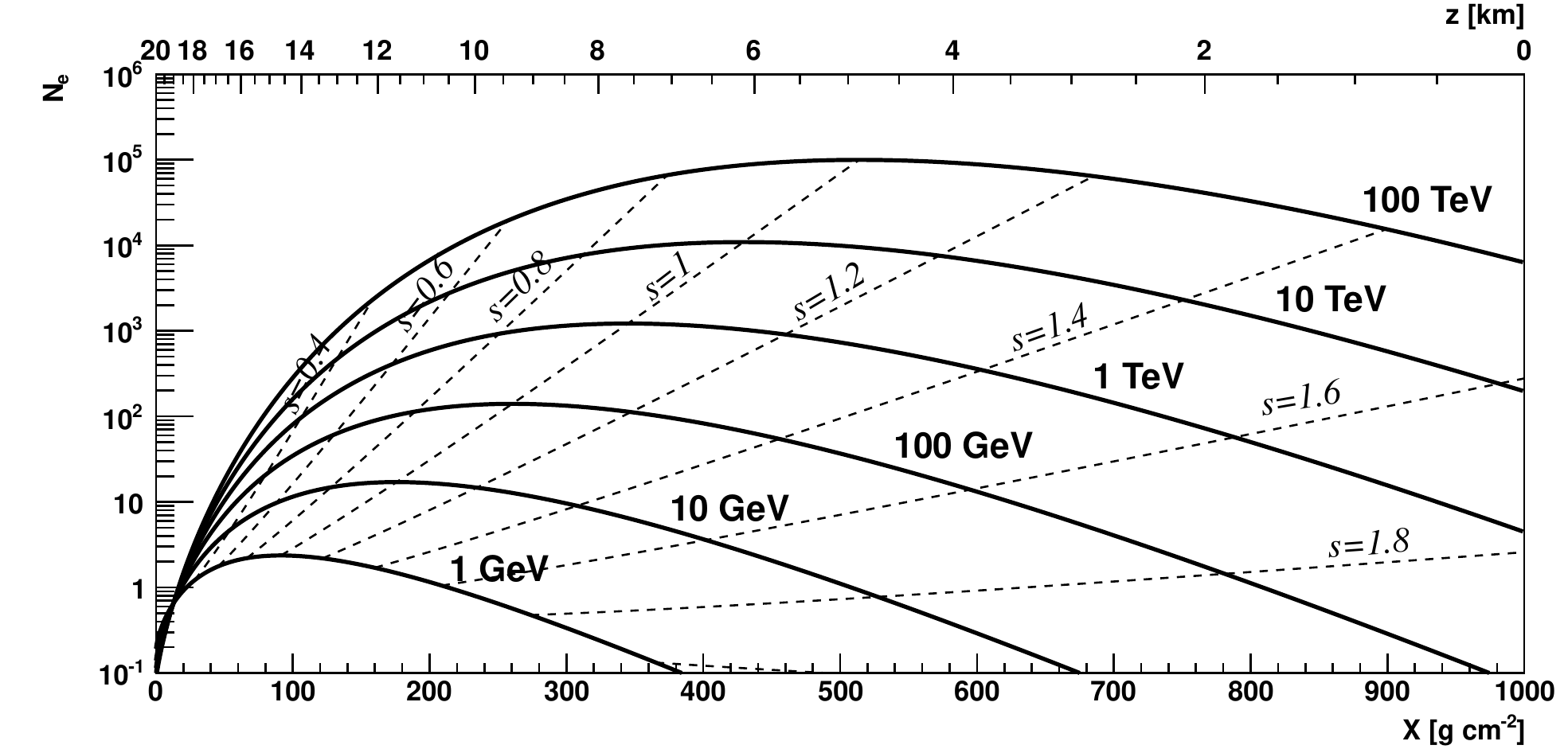} 
\end{center}
\caption{\label{fig:ShowerProfile}Illustration of the development of electromagnetic showers in an hydrostatic, isothermal atmosphere,
following the Greisen semi-empirical model. The solid lines indicate the number of electrons and positrons in the shower as function of depth.
The dashed lines corresponds to equal-age curves.}
\end{figure}

Orders of magnitude for the number of particles and the altitude of the maximum development of showers are given in Table~\ref{tab:Greisen}
and shown in Fig.~\ref{fig:ShowerProfile}.

\begin{table} [hb!]
\begin{center}
\begin{tabular}{|c|c|c|c|}
\hline
$E_{0}$ & $T_{\mathit{max}}(\U{g}\UU{cm}{-2})$ & Altitude (m) & $ N_{e}(t_{\mathit{max}})$ \\
\hline
$30\U{GeV}$ & $216$ & $12000$ & $50$\\
$1\U{TeV}$ & $345$ & $8000$ & $1200$\\
$1000\U{TeV}$ & $600$ & $4400$ & $0,9 \times 10^6$\\
$10^{19} \U{eV}$ & $936$ & $1200$ & $7,4 \times 10^9$\\
\hline
\end{tabular}
\caption{\label{tab:Greisen}Orders of magnitude of shower development for different primary energies}
\end{center}
\end{table}

\subsection{Cherenkov Radiation}

\begin{figure}[b!]
\centering
\includegraphics[width=0.45\textwidth]{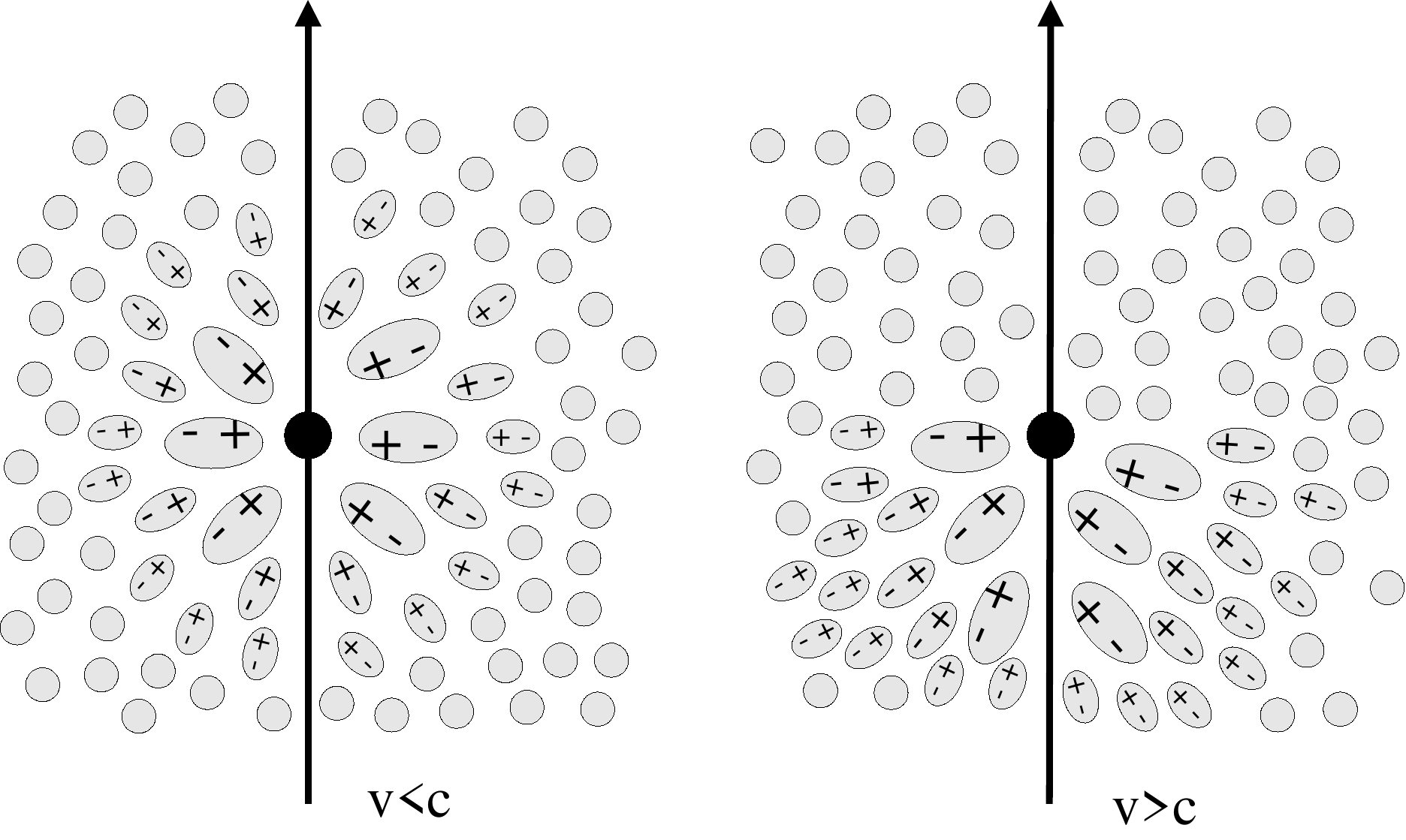}
\hfill
\includegraphics[width=0.45\textwidth]{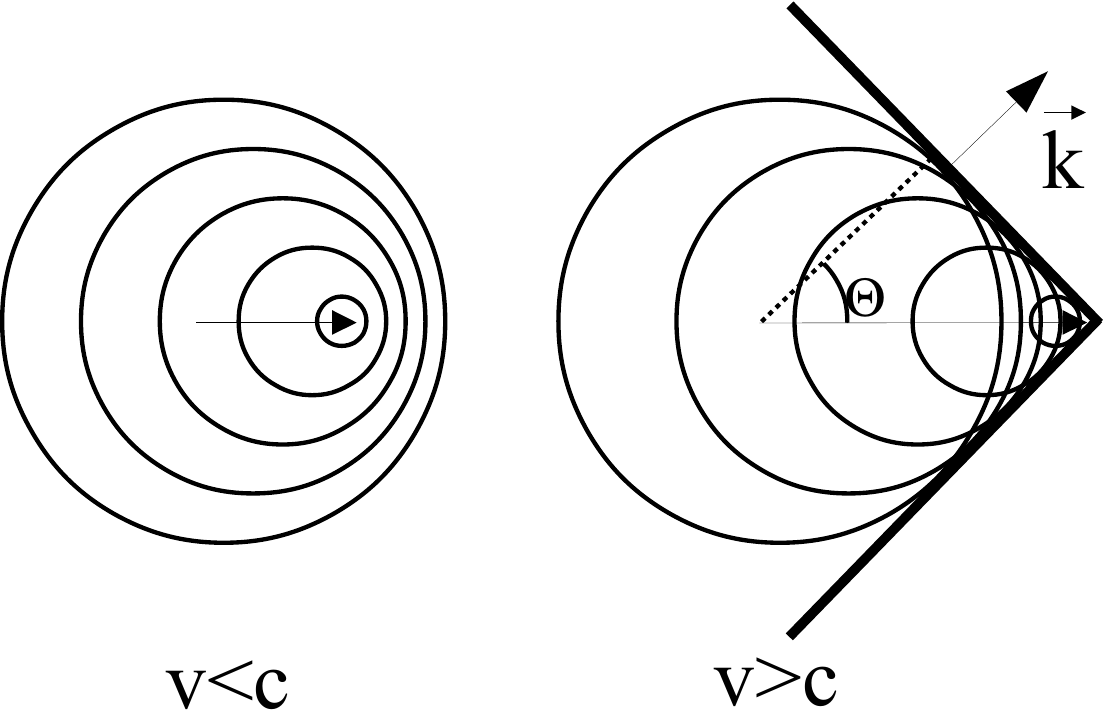}
\caption{\label{fig:Detection:CherenkovPolarisation}{\bf Left: }Illustration of the polarization of the medium induced by the crossing
of a relativistic particle. {\bf Right: }Construction of Cherenkov wave-front.}
\end{figure}

Ultra-relativistic particles in the shower travel faster than the speed of light in the air. Coherent depolarization of the dielectric
medium (of refractive index $n$) results in a forward-beamed emission called ``{\it Cherenkov Radiation}'', emitted along a cone with opening angle $\theta_c$,
and with a number of photons emitted per unit track length $\mathrm{d} z$ of charged particle and per unit wavelength $\mathrm{d}^2 N/\mathrm{d} z \mathrm{d} \lambda$
(See Fig~\ref{fig:Detection:CherenkovPolarisation}):

\begin{equation}
\cos \theta_c = \frac{1}{\beta n}, ~~~~~ \frac{\mathrm{d}^2 N_{ph}}{\mathrm{d} x \mathrm{d} \lambda} = 2 \pi \alpha Z^2 \frac{\sin^2 \theta_c} {\lambda^2}
\end{equation}

In general, the refractive index depends on the density of air (and therefore on the altitude), so the Cherenkov yield does as well. The refractive index
of air is mainly a function of the pressure (or density):

\begin{equation}
(n - 1) = 2.92 \times 10^{-4} \times \frac{P}{P_0} \times \frac{288.15 \U{K}}{T}
\end{equation}

In the simplified case of an hydrostatic, isothermal atmosphere, the density as function of altitude reads $\rho(z) = \rho_0 \exp (- z/z_0)$, with $z_0 = R T / g M = 8.4\U{km}$ and $\rho_0 = 1.2 \U{kg}\UU{m}{-3}$. 
Under the approximation of small angles, $\sin^2 \theta \approx 2 (n-1)$ and for essentially vertical charged particles, the
Cherenkov yield per unit {\it thickness} $\mathrm{d} X = X /z_0 \times \mathrm{d} z$, can be then expressed analytically:

\begin{equation}
\frac{\mathrm{d}^2 N_{ph}}{\mathrm{d} X \, \mathrm{d} \lambda} = 2 \pi \alpha Z^2 \frac{\sin^2 \theta_c} {\lambda^2} \times \frac{z_0}{X}
\approx  \frac {4 \pi \alpha Z^2 } {\rho_0 \lambda^2} \times 2.92 \times 10^{-4} \times \frac{288.15 \U{K}}{T}
\end{equation}

Remarkably, this quantity does not depend on the altitude: the Cherenkov yield, when expressed in the natural variable describing the shower development, does not
depend on the local density. Under these approximations, the total amount of Cherenkov light emitted by a shower is given by an integral over the shower age:

\begin{equation}
\frac{\mathrm{d} N_{ph}}{\mathrm{d} \lambda} = 
\int \mathrm{d} X N_e(t) \times \frac{\mathrm{d}^2 N_{ph}}{\mathrm{d} X \, \mathrm{d} \lambda}
= 6 \times 0.31 \sqrt y X_0 \times \frac{E_0}{E_c} \times \frac{\mathrm{d}^2 N_{ph}}{\mathrm{d} X \, \mathrm{d} \lambda} \int \frac{\mathrm{d} s}{(3- s)^2} \exp \left( \frac{2 s}{3 - s} \left( 1 - \frac 3 2 \ln s \right) \right)
\end{equation}

To a correction factor $\sqrt{y}$ that varies only between 2.2 and 3.4 for $E_0$ between $10\U{GeV}$ and $10\U{TeV}$, the total amount of Cherenkov light is therefore
almost proportional to the primary energy, thus making a calorimetric measurement possible even in an strongly inhomogeneous environment.
This is further confirmed by more elaborated, realistic simulations.

\subsection{Angular distribution and light pool}

\begin{figure}[b!]
\centering
\begin{minipage}[c]{0.4\textwidth}
  \includegraphics[width=\textwidth]{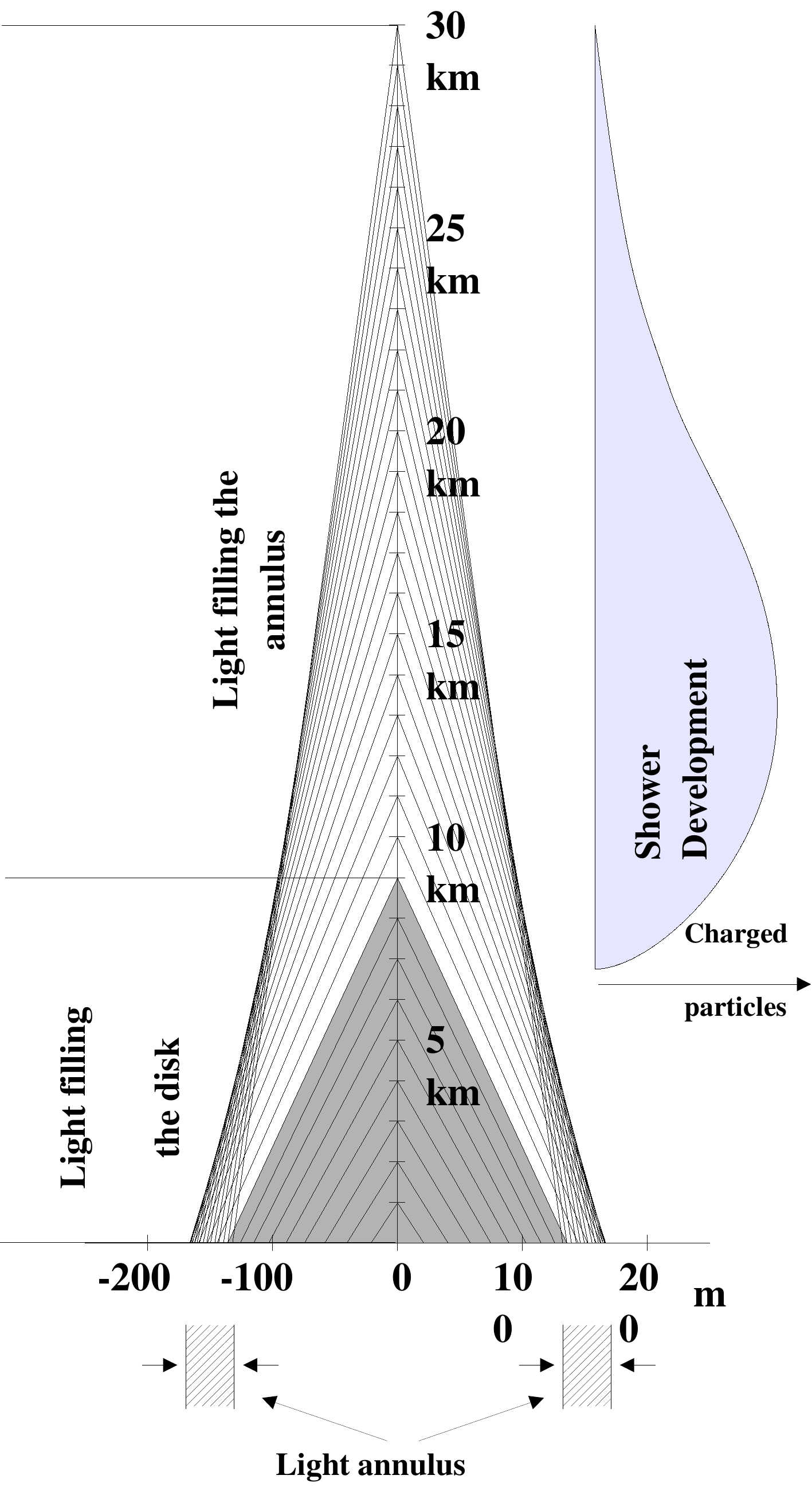}
\end{minipage} \hfill
\begin{minipage}[c]{0.55\textwidth}
  \includegraphics[width=\textwidth]{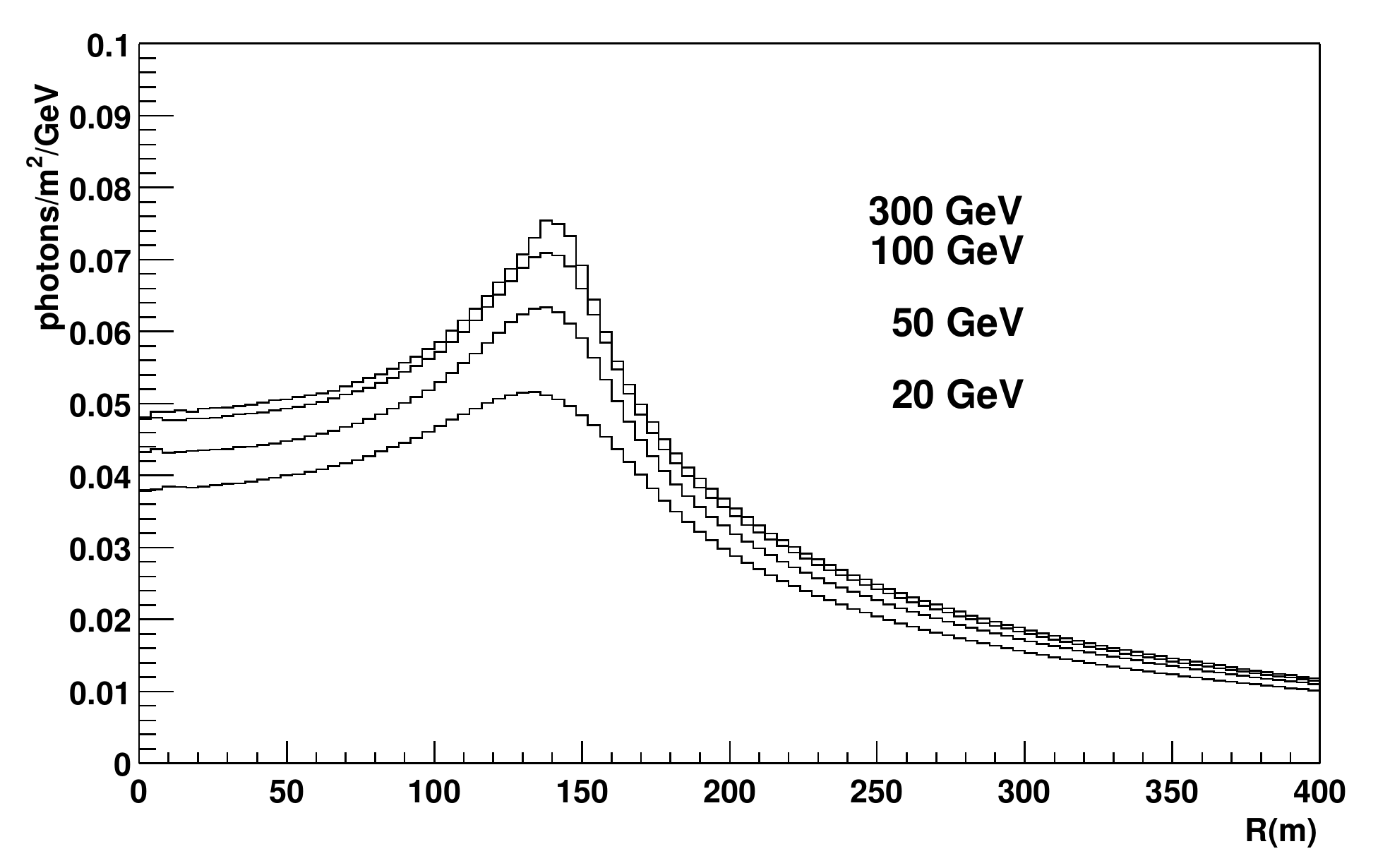} \\ \vfill
  \includegraphics[width=\textwidth]{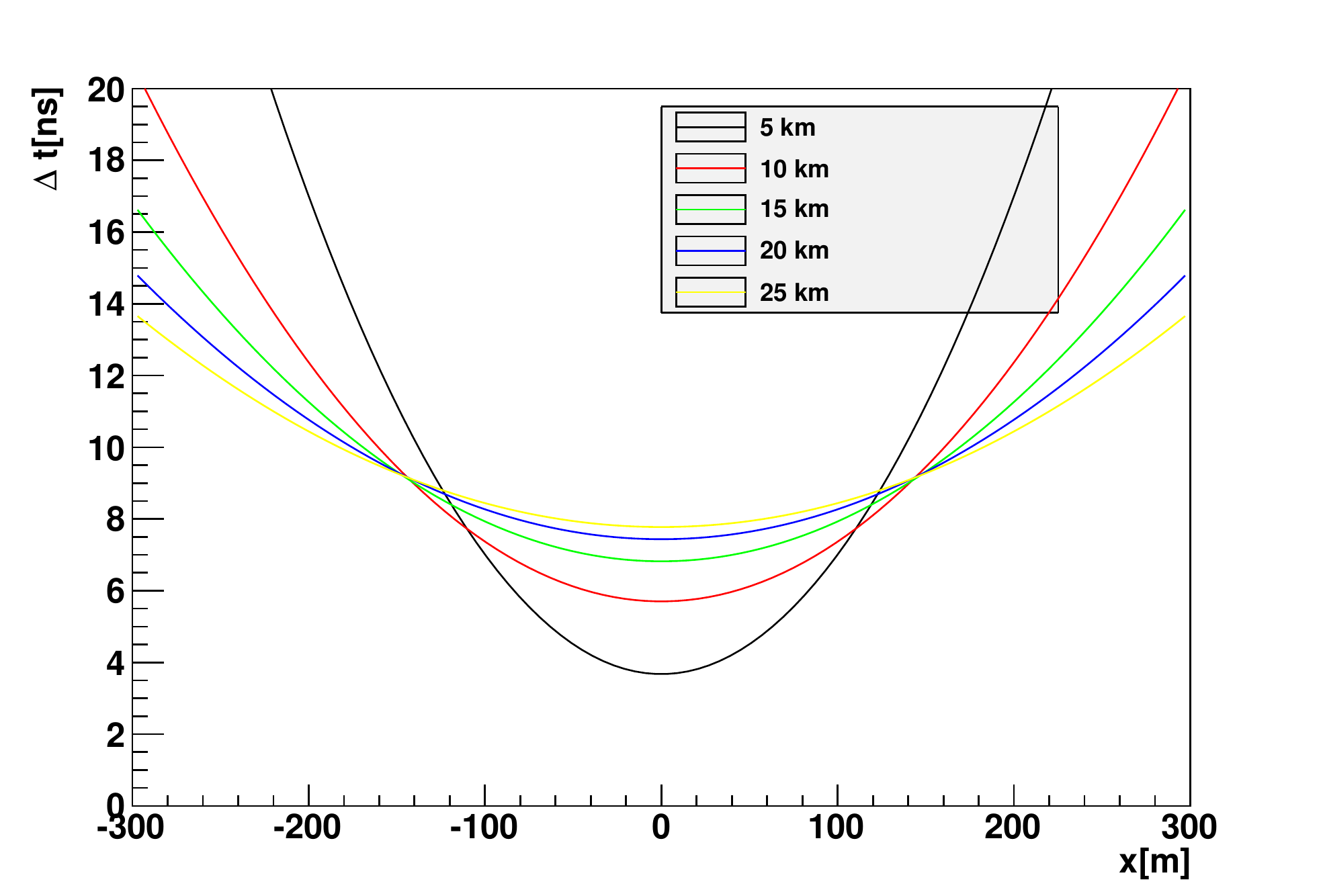}  
\end{minipage}
\caption{\label{fig:Shower:ShowerDevelopment}{\bf Left:} Shower development. {\bf Top Right:} Lateral profile of showers of different energies at sea level. {\bf Bottom Right:} Time delay as function of lateral distance for various altitudes of
emission. }
\end{figure}

Due to the evolution of the atmospheric density with altitude, the
Cherenkov angle increases from $\sim 0.2^\circ$ at an altitude of $\sim 30\U{km}$ to $\sim 1.5^\circ$ at sea level.
The effect of this variation is illustrated for vertical showers in Fig.~\ref{fig:Shower:ShowerDevelopment}, left, and
is responsible for the formation of a light annulus at a distance of $\sim 150\U{m}$ from
the shower impact on the ground: the variation of the Cherenkov angle with altitude almost
exactly compensates the effect of the varying distance to the ground.

Similarly, the spread of the arrival time of the photons on the ground results from two different effects, somewhat in competition:
The charged particle in the shower travels faster than light. Therefore, close to the shower axis, the photons emitted at low
altitude reach the detector {\it before} those emitted at high altitude. At large impact distance however, the photons emitted
at low altitude have a longer geometrical trajectory (track of the charged particle to the emission point + track of the photon
itself) than those emitted at high altitude, and reach the detector {\it after} the latter. 
At a distance of $\sim 120\U{m}$ from the shower (Fig.~\ref{fig:Shower:ShowerDevelopment}, bottom right), the two effects compensate
almost exactly, resulting in a very short duration of the shower of $\sim 2\U{ns}$. The shower duration can reach $\sim 5 \U{ns}$ on axis,
and increases significantly for impact distances $> 200\U{m}$. The time integration window of the detectors therefore has
a direct impact on their effective area (through their capability to detect distant showers) but also on the amount of integrated
night sky background light and therefore on their energy threshold.

\subsection{Detection Techniques}

VHE gamma-ray astronomy rests on two basic detector technologies:
\begin{itemize}
\item Detectors that measure particles of the shower tail reaching the ground. 
This method provides a snapshot of the shower at the moment it reaches the ground and constitutes 
 the so-called ``{\it particle sampler}'' technique.
Those detectors have a very large duty cycle (potentially $100\%$), but rather high energy threshold (as high energy showers are more penetrating and produce charged
particles at lower altitude than lower energy showers). Moreover, as they only have access to shower tails, they usually
have a rather poor capability to discriminate the showers induced by \grays\ from the much more numerous showers induced by protons and
charged nuclei. Such detectors are usually installed at high altitude to collect more charged particles.
Several types of particle samplers have been tried, including scintillator arrays~\cite{1999ApJ...525L..93A}, 
resistive plate chamber carpets~\cite{argo-2006} and water Cherenkov
ponds~\cite{2004ApJ...608..680A,2012APh....35..641A,2013APh....50...26A}. A sketch of the Milagro water Cherenkov detector is shown in Fig.~\ref{fig:Milagro}.
\begin{figure}[ht!]
\begin{center}
\includegraphics[width=0.8\textwidth]{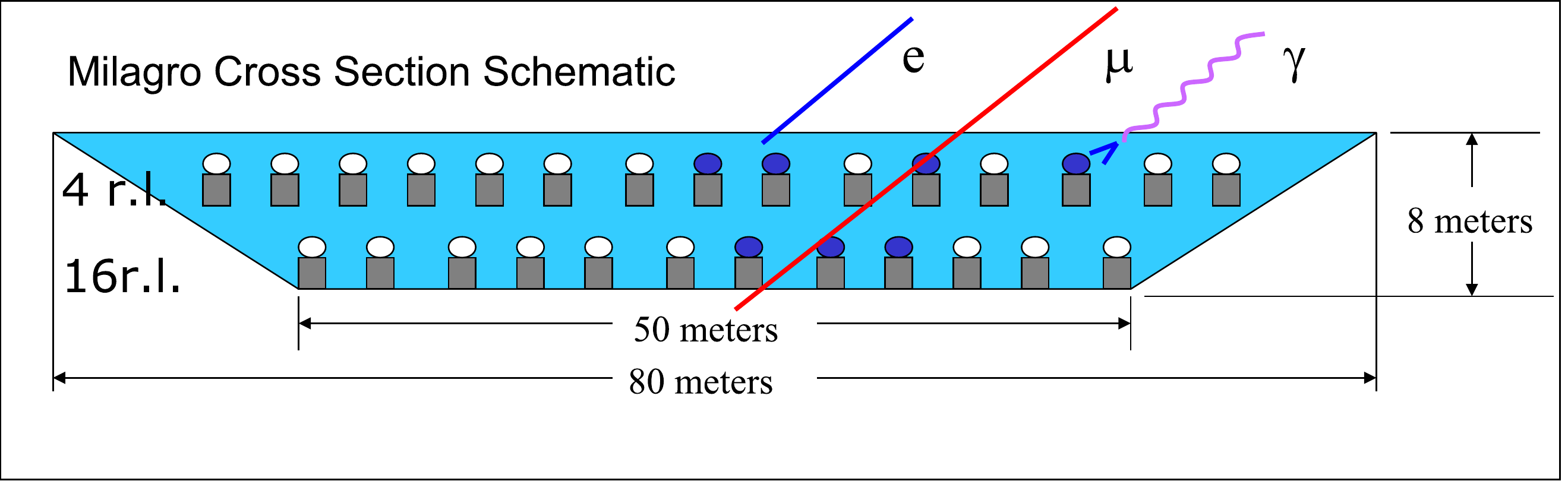}
\end{center}
\caption{\label{fig:Milagro}Sketch of the central Milagro water pond detector.}
\end{figure}
In the sampling technique, the direction reconstruction is based on the timing information (simple trigonometric
direction measurement using the time of arrival of the signal in each detector), completed by the spatial distribution
of the signal on the ground (mainly being used to determine the impact of the shower on the ground).

\item Cherenkov detectors for observing showers that died before reaching the ground, 
through the detection of the Cherenkov light produced in the atmosphere. 
This method uses the atmosphere as a calorimeter as described in the previous section. 
Several techniques have been tried in the past.
The most successful has been to use optical telescopes to take a ``picture'' of the showers (recording the Cherenkov light emitted by them), 
 as illustrated in Fig.~\ref{fig:Imageur}.
 These so-called ``{\it imaging atmospheric Cherenkov telescopes}'' (IACTs) are characterized by a relatively small field of view
(a few degrees angular diameter), low duty cycle ($\sim 10\%$, corresponding to moonless, clear nights), but a very large effective area, corresponding
to the size of the light-pool illuminated by the showers ($\sim 10^5\UU{m}{2}$), and very powerful discrimination capabilities. 
The experimental challenges are, on the one hand,
the very low intensity and short duration of the signal, requiring very fast and sensitive acquisition systems, and,
on the other hand, the huge background, from both the night sky luminosity and from the air showers produced by
charged cosmic rays.
The optimal altitude ($\sim 1500 \U{m}$) for this technique results from a trade-off between the transparency 
of the atmosphere to Cherenkov light (pushing for higher altitude) and the development of the shower (leading to larger effective areas, better shower containement and
improved calorimetric capabilities\footnote{To make a calorimetric measurement possible, the shower must die before reaching the ground 
so that the Cherenkov emission reflects the number of charged particles in the shower.} at low altitude).

\begin{figure}[t!]
\begin{center}
\includegraphics[width=0.48\textwidth]{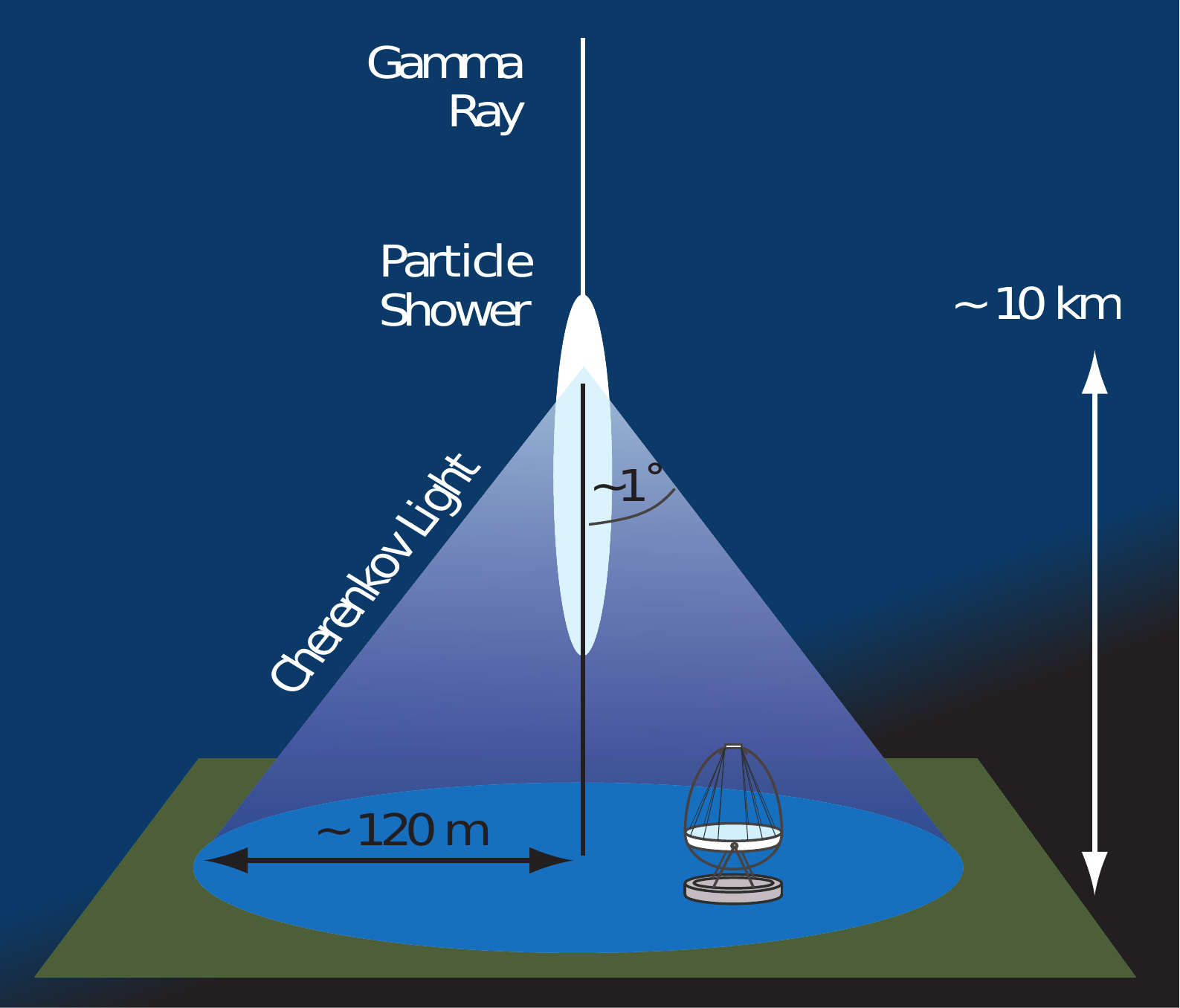} \hfill  
\includegraphics[width=0.35\textwidth]{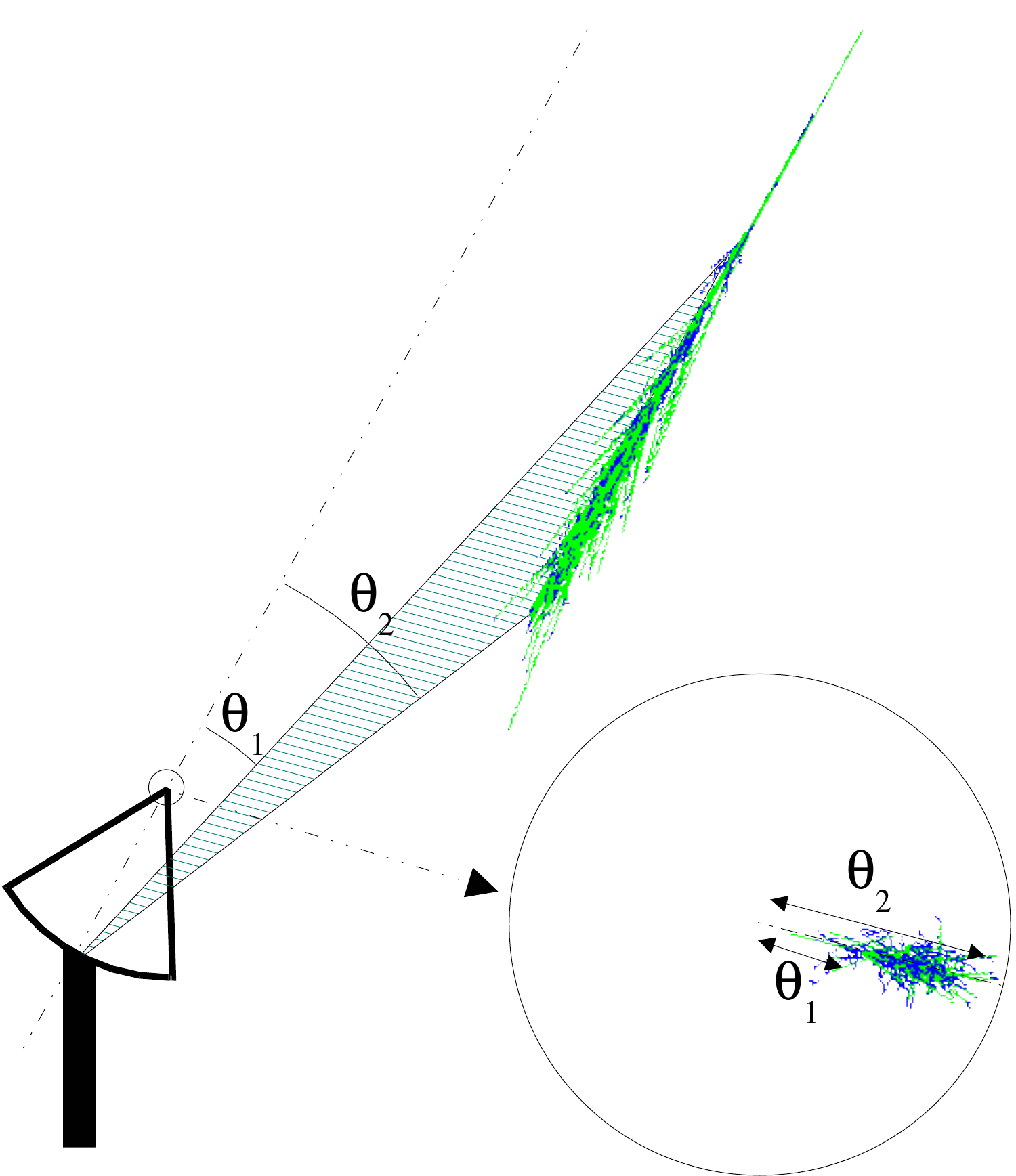}
\end{center}
\caption{\label{fig:Imageur}Imaging Atmospheric Cherenkov technique. {\bf Left:} the Cherenkov light emitted by the charged
particles in the shower is collected by several dishes. {\bf Right:} The shower angular image is projected into the
camera focal plane.}
\end{figure}
\end{itemize}


\section{The pioneering era}
\label{sec:pionners}

\subsection{Early days}

After the discovery of cosmic rays by V.~Hess, many experiments were designed to 
study their nature and origin with improved detection techniques.
In the 1920s it was a popular belief that all cosmic rays originated from gamma rays. The debate between Robert Millikan (arguing that
electrons reaching the earth were produced by Compton scattering of high energy gamma-rays) and
Arthur Compton (claiming that cosmic rays were genuine charged particles) made the cover page of the New York Times in 1932.
It took 27 years after Hess' first discovery of CRs until Pierre Auger discovered extended air showers initiated by CRs  hitting the
atmosphere~\cite{Auger1939a}. The understanding of the shower process grew with time and cosmic ray physicists
built balloons to study low energy charged CRs and air shower arrays to detect the high energy tail of the CR spectrum.
Meanwhile it was becoming clear that gamma rays are only a tiny fraction of the ionizing radiation hitting the atmosphere.
In the early 1980s it was mostly thought that about $1\%$ of the CRs were gamma rays. Nowadays this question is still not completely 
solved and much smaller flux ratios are assumed. 
One currently estimates that at most $10^{-4}$ of all particles coming from the Galactic plane are gamma rays,
and that an even smaller fraction ($\sim 10^{-5}$) of particles from outside the galactic plane are gamma rays.
The expectation in the early days of high gamma-rays fluxes led to strong enthusiasm about abilities to detect
sources of gamma rays, but with time these expectations vanished as no unambiguous detection of gamma-ray
signals was successful.

It took 19 years to detect Cherenkov light from air showers after the discovery of the effect~\cite{Cherenkov1934a}.
In 1953, W.~Galbraith and J.~V.~Jelley built a simple detector and proved that air showers do actually generate Cherenkov light, 
which could be detected as a fast light flash during clear dark nights~\cite{Galbraith1953a}. 
With a threshold of around four times the night sky-noise level, they observed signals with a rate of about one event every two to three
minutes.
The early detectors consisted of a very simple arrangement, mainly a search-light mirror viewed by a single 
photomultiplier tube (PMT) as light detector.
The first setup was installed in a garbage can for shielding from stray light. 
In the following years, the technique was refined by using larger
mirrors, replacing the single PMT by a few ones arranged in the focal plane and 
even by trying to detect coincidences between several such simple telescopes .

The three decades from 1960 to the end of the 1980's saw steady but rather slow progress towards discovering
sources of VHE gamma rays. Experiments provided doubtful and often inconsistent
results and the funding agencies were not willing to fund large installations, which led many physicists
to leave the field.

In 1977 T.C.~Weekes, in collaboration with K.E.~Turver, presented for the first time
concepts for the separation of gamma-ray initiated showers from hadronic background showers~\cite{1977ESASP.124..279W}.
In this work, where computer simulations were used for the first time, the advantages
of stereoscopic observation was already advocated.

The ``{\it imaging technique}'', first proposed by A.M.~Hillas in 1985~\cite{Hillas1985} and developed during these days,
consisted of placing a fast camera (initially of 37 PMT's) at the focal plane of a telescope (Fig.~\ref{fig:Imageur}) in order to record the image of the shower
with an angular resolution better than its size.
This technique quickly looked the most promising, which encouraged a few dedicated enthusiasts to stay 
and pursue developing the instrumentation and the data analysis techniques.
Finally, in 1989 a first gamma-ray source, the Crab Nebula, was unambiguously detected 
with more than 9 standard deviations
by the Whipple telescope under the lead of Trevor Weekes~\cite{whipple-crab}. It was the culmination of a long journey for the group that had
started in 1968 with a large $10\U{m}$ telescope that was completed at the Fred Lawrence
Whipple Observatory on Mount Hopkins in Arizona, USA~\cite{Fazio1968a}. 


Three major factors led to the detection of the first gamma-ray source~\cite{Lorenz2012a}:
\begin{itemize}
 \item The Whipple collaboration focused on a source that turned out to be the strongest 
steady state gamma-ray emitter: the Crab Nebula. 
 \item They used a large light collection area telescope ($10\U{m}$ diameter)
and an imaging camera with 37 PMTs covering a field of view of 3.5 degrees. 
This allowed recording of true images of air showers, which could be used 
for an efficient gamma/hadron separation which was not possible with telescopes
 without imaging cameras.
 \item The team invested a lot of effort to improve the analysis and gamma/hadron
discrimination methods. The analysis developed by the Whipple collaboration
in the mid-eighties was based on the combination of 
a measurement of the shower image orientation~\cite{Weekes1983a} and 
on differences in image shapes between gamma-ray- and hadron-induced showers~\cite{Stepanian1983a}. 
The originally rather simple analysis, based on image first and second moments\footnote{Deriving the images moments 
is equivalent to modeling that image by a two-dimensional ellipse in the camera, as described later
in section \ref{sec:Digging}.}
commonly known as Hillas parameters~\cite{Hillas1985},
became the basic concept for gamma/hadron separation in the following Cherenkov telescope
experiments.
\end{itemize}

\subsection{In all directions ...}


Subsequent (and mostly contemporaneous with Whipple) experiments confirmed the gamma-ray signal from the Crab Nebula:
Crimean GT48 Observatory~\cite{Vladimirsky1989a,Fomin1991a},
Yerevan~\cite{Aharonian-Yerevan1989a}, 
Ala-Too~\cite{Nikolsky1989a}, 
Cangaroo-I~\cite{Kifune1992a},
The HEGRA array~\cite{Aharonian-HEGRA-1991a},
Granite (Whipple+$11\U{-m}$ Tel.)~\cite{Akerlof1990a},
MarK~V~\cite{Bowden1991a}, 
ASGAT~\cite{Goret1993}, Themistocle~\cite{Themistocle1993},
Telescope Array prototype (TA coll.)~\cite{Aiso1997a} and
CAT~\cite{Barrau1998a}.
In 1992, the Whipple collaboration made another great discovery: a signal from an active galactic nuclei (the blazar Markarian~421, located
at a red-shift of $z=0.031$)
was detected on a level of $30\%$ of the the signal from the Crab Nebula~\cite{whipple-421}.
This opened the window of extragalactic gamma-ray searches.

In 1992, Patrick Fleury and Giuseppe Vacanti invited the community to a conference at
Palaiseau with the aim of forming a project of a major imaging Cherenkov telescope~\cite{Fleury1992a}.
Despite a large participation at the conference and a strong excitement of the community about the new discoveries
it was not possible to converge on a joint large scale project. Instead, individual groups
pursued developments of the Cherenkov technique in different directions.

Major ideas and experiments followed in the 1990s, exploring different instrumental paths in parallel.

\subsubsection{Developments in Imaging Atmospheric Cherenkov Telescopes}

\begin{itemize}
 \item The Cherenkov Array at Th\'emis (CAT) collaboration built a single dish Cherenkov telescope 
with $4.7\U{m}$ diameter~\cite{Barrau1998a}. The novelty of the telescope was a fine grained camera consisting of
546 pixels with an angular size of $0.12^\circ$ each.
The use of such a camera allowed an accurate analysis of the longitudinal
and lateral light profile of the shower image, as discussed in~\cite{LeBohec1998}, giving a good separation
of gamma-ray showers from hadronic ones by means of a goodness of the fit variable 
and of the pointing angle $\alpha$.
 \item The HEGRA array in the Canaries island, consisting of 5 IACTs each having $3.3\U{m}$ diameter mirror reflector and a 
camera of 271 pixels, made of PMTs, demonstrated the power of stereoscopic observations.
The telescopes were arranged on a square of $100\U{m}$ side length with an additional telescope centered in the square.
HEGRA did not actually pioneer stereoscopic observations based on a telescope coincidence trigger -- this was already done by the Crimean
observatory~\cite{Chudakov1963a} -- but it was the first one to find the right separation between the telescopes
thanks to Monte-Carlo simulations.
Multiple shower images provided information for the optimum discrimination between the Cherenkov light flashes of 
gamma-ray and cosmic-ray induced showers. Based on the stereo views, 
an unambiguous reconstruction of the air showers in space became possible, 
leading to a very significantly improved angular reconstruction of primary particles on an event-by-event basis
as well as a much stronger rejection of cosmic-ray induced showers.
 \item The CANGAROO collaboration constructed a Whipple-like telescope in Australia.
 \item A group from Durham University upgraded their Mark-VI  telescope in Narrabri, Australia 
combining an idea of a 3-fold coincidence between signals from 
reflector dishes placed next to each other (the idea originally developed and implemented
with smaller reflectors by the Crimean observatory in the 1960s~\cite{Chudakov1963a})
and an imaging camera in the focus of each dish~\cite{Bowden1991a}. 
The Mark-VI telescope was successfully operated between 1995 and 2000.
\end{itemize}

\subsubsection{Wave-front sampling technique}

\begin{itemize}
\item The ASGAT~\cite{Goret1993} and Themistocle~\cite{Themistocle1993} arrays,
installed in 1988 on the site of a former solar power plant, ''Themis'', in the french Pyrenees,
were using a completely independent technique based on fast-timing wave front sampling.
ASGAT used 7 parabolic dishes of $7\U{m}$ diameter arranged on an hexagonal shape, whereas 
Themistocle was based on 18 much smaller mirrors ($80\U{cm}$ diameter) but spread on a larger area ($170\U{m}\times 300\U{m}$).
Both experiments proved that timing information was useful for reconstructing the shower parameters and for distinguishing
gamma rays from charged cosmic rays, quickly confirmed the detection of the Crab Nebula and extended the measurement of its energy 
spectrum well above $1\U{TeV}$.
 \item Following this first success in the wave-front sampling technique, several groups tried to convert solar 
farms into gamma-ray detectors during night time.
These solar farms were based on the concept of fields of heliostats reflecting the light of the sun
onto a furnace at the top of a $\sim 100\U{m}$ tower. The use of many large-area
heliostats collecting Cherenkov light and focusing it onto a single detector mounted on a tower had advantages 
to save costs and achieve a low energy threshold. Four installations were in operation: 
CELESTE in the French Pyrenees~\cite{Pare2002a}, STACEE detector using a
prototype solar power station near Albuquerque, New Mexico~\cite{Gingrich2005a},
the Keck Solar Two in Barstow, California~\cite{Tuermer1999a}, and the GRAAL detector 
using the heliostats of the Plataforma Solar in Almer\'ia, Spain~\cite{Plaga199a}.
Although the approached proved to be very successful in terms of reduction of the energy threshold, the very narrow
field of view of these instruments ($\sim 0.5^\circ$, corresponding to the angular size of the sun)
resulted in a very challenging background subtraction and event identification: the detectors were actually
only observing a fraction of the hadronic showers, making them resemble electromagnetic ones.
All detectors based on solar heliostat plants were able to detect the strongest sources like the Crab
nebula and the flaring Mkn 421 and Mkn 501 but did not reach the sensitivity of the third-generation
imaging telescopes and therefore the method was abandoned around 2005~\cite{Lorenz2012a}.
\end{itemize}

\subsubsection{Particle samplers}

\begin{figure}[ht!]
\begin{center}
\includegraphics[width=0.9\textwidth]{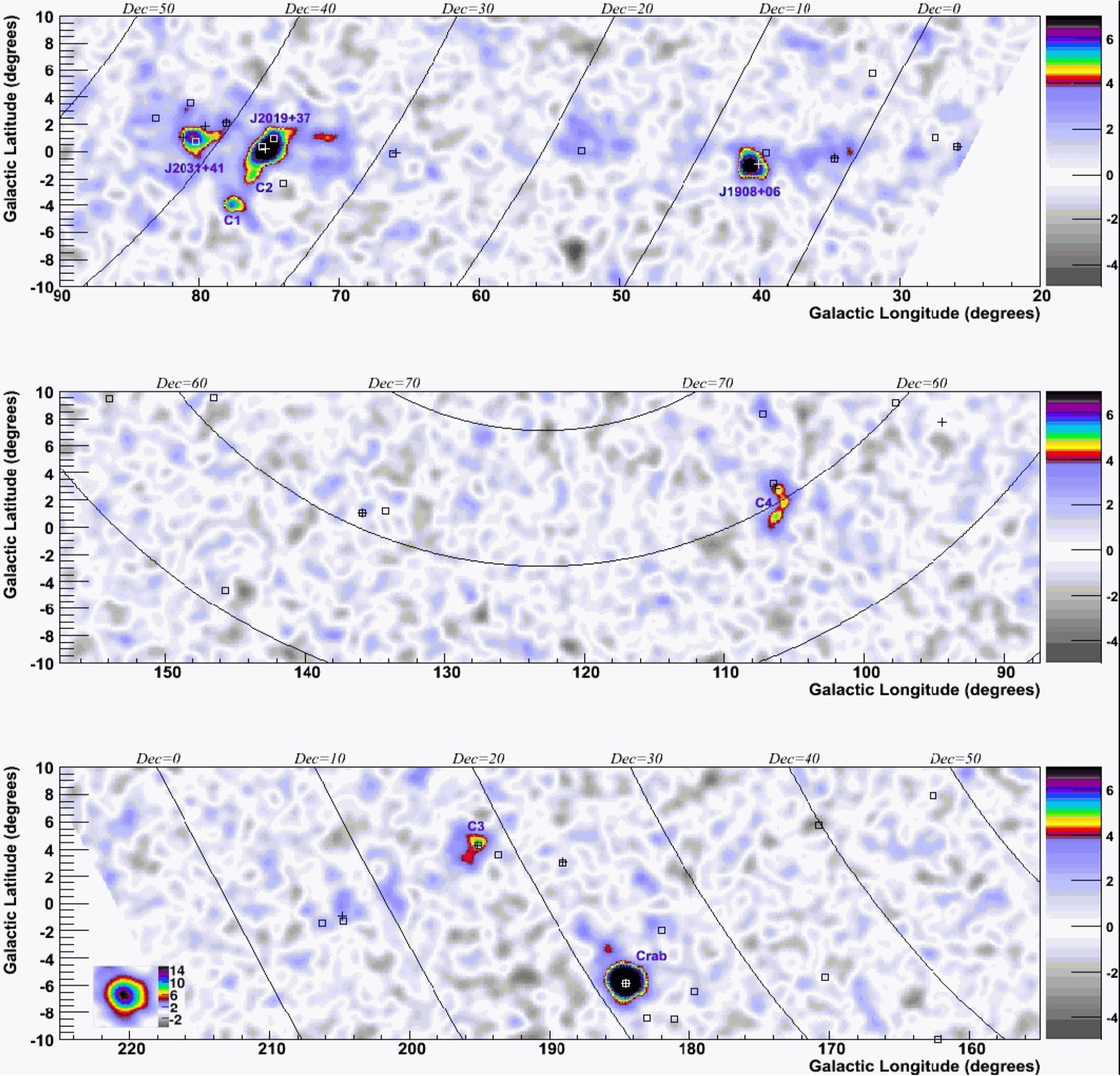}
\caption{\label{fig:MilagroDiffuse}Significance map of the galactic emission as seen by the Milagro Water Cherenkov experiment.
The color code shows the pretrial significance, smoothed according to the instrument PSF. From~\cite{milagro-2007}.}
\end{center}
\end{figure}

\begin{itemize}
 \item After a first successful prototype Water Cherenkov detector (MILAGRITO~\cite{Milagrito:2000NIMPA.449..478A}),
the Milagro~\cite{2004ApJ...608..680A} experiment operated between 1999 and 2008 in a former water reservoir in the Jemez mountains near Los Alamos, 
New Mexico, at an altitude of $2630\U{m}$.
Milagro (see Fig.~\ref{fig:Milagro}) was composed of a central $60\U{m} \times 80 \U{m}$ pond completed by a sparse $200\U{m}\times 200 \U{m}$ array
of 175 {\it ``outrigger''} water tanks surrounding it. The pond was instrumented with 723 photo-multiplier tubes arranged in two layers
the top one being dedicated to the measurement of the electromagnetic component in showers, and the
bottom one, located $6\U{m}$ below the surface, dedicated to the identification of hadronic showers through their
muonic content.  Milagro's large field of view ($2\U{sr}$) and high duty cycle ($>90\%$) allowed it
to monitor the entire overhead sky continuously, making it well suited to measuring diffuse emission.
 This is illustrated by the view of the Galactic Plane by Milagro~(Fig.~\ref{fig:MilagroDiffuse}) above $\sim 20\U{TeV}$ ,
representing 2358 days of data collected by Milagro between July 2000 and January 2007.
\end{itemize}

At the end of this era of so-called second generation of Cherenkov telescopes (around the year 2000) 
7 gamma-ray emitters were established:
The Crab pulsar wind nebula\footnote{A pulsar wind nebula is a synchrotron nebula, confined by the reverse 
shock of an expanding supernova shell, and fed by an energetic pulsar.}
and the supernova remnant RX~J1713.7-3946 as Galactic emitters~\cite{CRAS_PWN} and
Mrk~421, Mrk~501, 1ES~1959+650, PKS~2155-304, and 1ES\,1426+428 as extragalactic sources~\cite{CRAS_EGAL}, with 
1ES\,1426+428 being the most distant one (redshift $z=0.129$).

\subsection{\label{sec:Digging}Digging in the data}

The advent of more powerful computers allowed the development of
the showers to be simulated to a much better precision, and opened the way to more efficient 
and elaborated analysis techniques, able to take full advantage of the
fine-grained camera with improved acquisition speed.

From the beginning of ground-based gamma-ray astronomy, data analysis 
techniques have been mostly based on the ``{\it Hillas parametrization}''~\cite{Hillas1985}
of shower images, relying on the fact that  gamma-ray images in the focal plane are, to 
a good approximation, elliptical in shape and intrinsically narrower than hadronic images. In 1985, based on pioneering Monte Carlo simulations, 
A.M.~Hillas proposed to reduce the recorded images to a few parameters, constructed from the first and second moments of the 
light distribution in the camera, and corresponding to the modeling of the image by a two-dimensional ellipse. 

\begin{wrapfigure}[12]{r}{6cm}
\includegraphics[width=4cm]{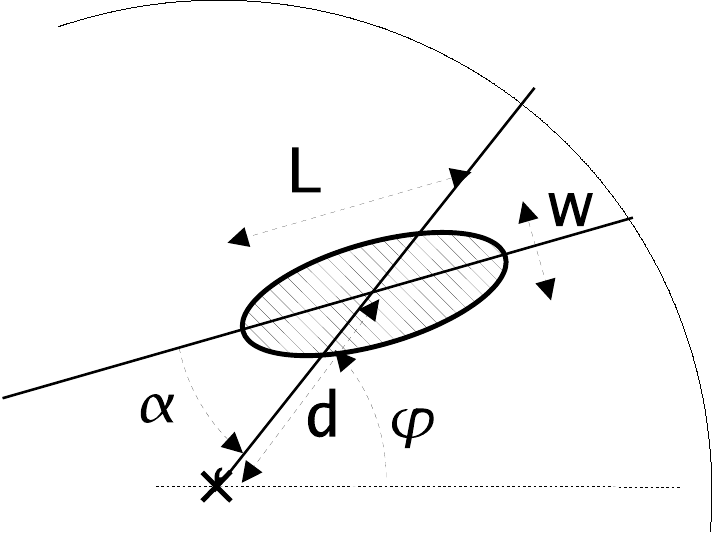}
\caption{Geometrical definition of the Hillas Parameters. From~\cite{HDRMathieu}.}
\label{fig:Hillas}
\end{wrapfigure}

These parameters, shown in Fig.~\ref{fig:Hillas},
are the following:

\begin{itemize}
\item image center of gravity (first moments)
\item length $L$ and width $w$ of the ellipse (second moments)
\item size (total charge of photo-electrons in the image)
\item nominal distance $d$ (angular distance between the centre of the camera
and the image centre of gravity)
\item azimuthal angle of the image main axis $\phi$ (second moments)
\item orientation angle $\alpha$ (see Fig.~\ref{fig:Hillas}.)
\end{itemize}

\begin{figure}[htbp!]
\begin{center}
\begin{tabular}{ccc}
\includegraphics[width=0.35\textwidth]{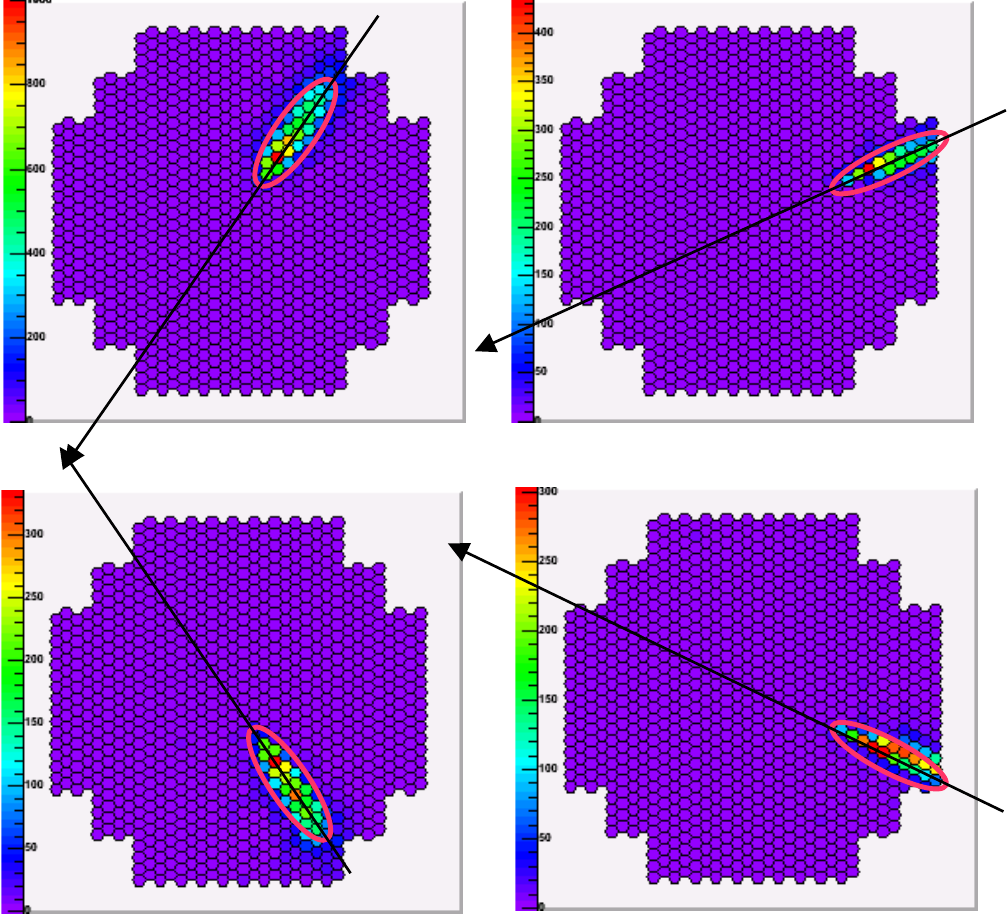} &
\includegraphics[width=0.35\textwidth]{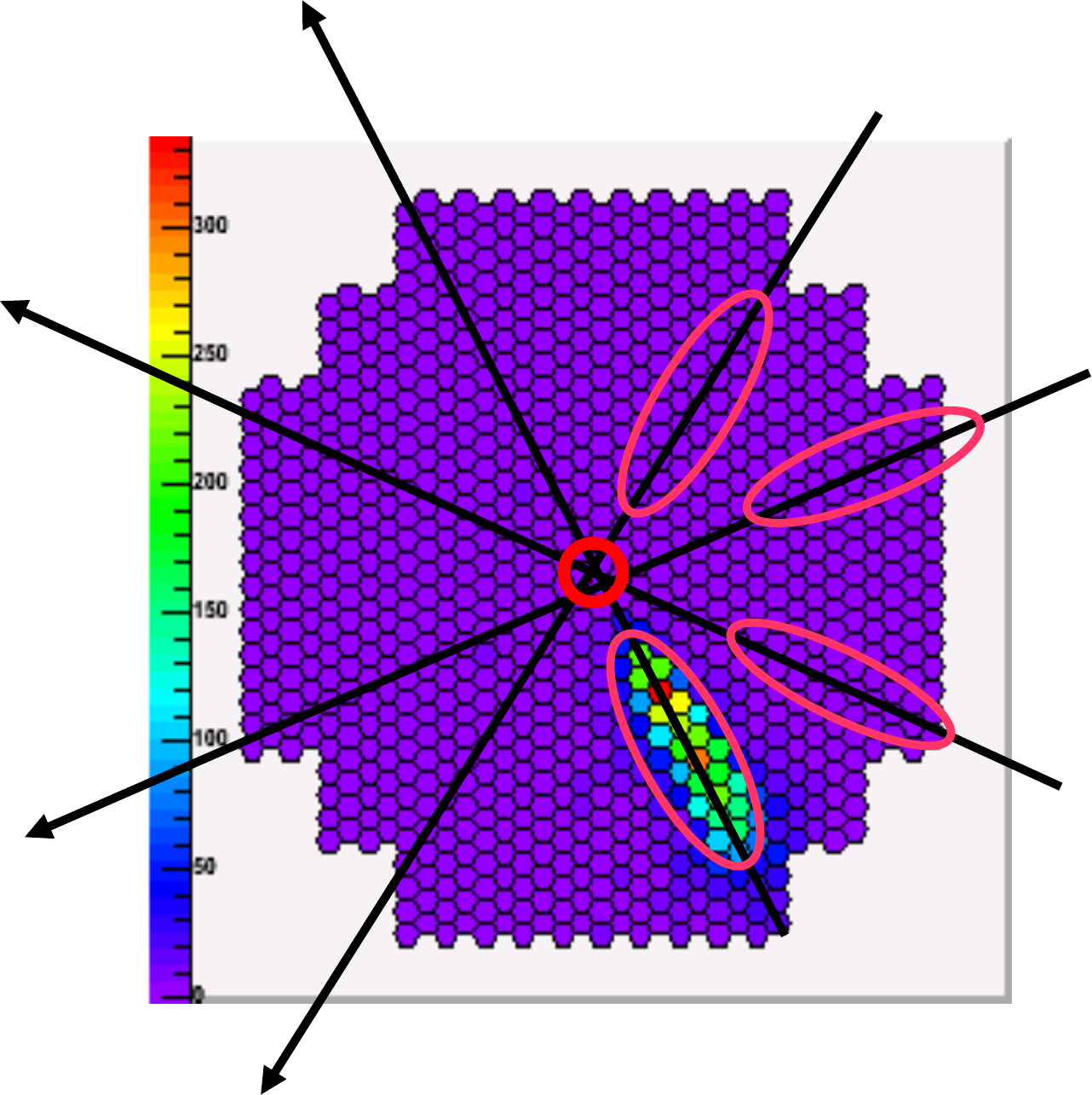} &
\includegraphics[width=0.26\textwidth]{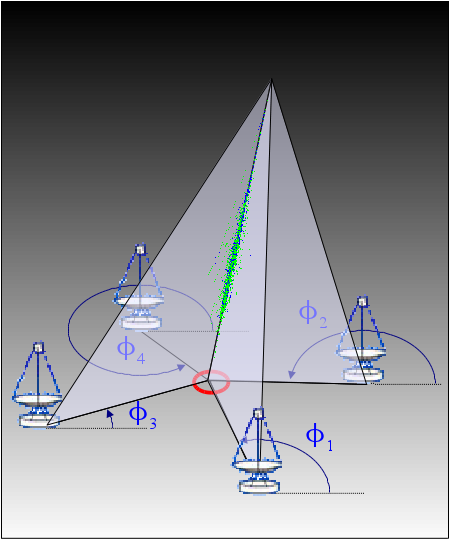} \cr
\end{tabular}
\caption{Geometric reconstruction of shower direction and impact in stereoscopic mode. {\bf Left:} In the camera frame, the main axis of the shower corresponds
to a plane that contains the actual shower track and the telescope. The primary particle direction corresponds to a point on this
main axis. {\bf Middle:} The intersection of the main axis of the images recorded by the different telescopes immediately provides
the primary particle direction. {\bf Right:} Direct intersection of the planes containing the shower tracks and the telescopes
provides the shower impact on the ground. From~\cite{HDRMathieu}.}
\label{fig:HillasStereoDirection}
\end{center}
\end{figure}

The stereoscopic imaging technique, already advocated in 1977~\cite{1977ESASP.124..279W} and successfully developed by HEGRA~\cite{Daum1997} in 1997, allowed a simple, 
geometrical reconstruction of the shower direction and impact parameter and resulted in a major step
in angular resolution as well as in background rejection. The source direction is given by the intersection 
of the major axes of the shower images in the camera (Fig.~\ref{fig:HillasStereoDirection}), and the shower impact point is obtained
in a similar manner, using a geometrical intersection of the planes containing the telescopes and
the shower axes. The energy is then estimated from a weighted average of each single
telescope energy reconstruction. The separation between the showers induced by gamma rays and those induced by charged cosmic rays originates
mainly from the larger width of the later.

The Hillas parameters not only allow the reconstruction of the shower parameters, but also provide
some discrimination between \gray\ candidates and the much more numerous hadrons, based on the extension (width and length) of the recorded images.
Several techniques have been developed, exploiting to an increased extent the existing correlation
between the different parameters~\cite{Reynolds1993,Daum1997,Mohanty1998}.
%
As an example, in the so-called {\it Scaled Cuts} technique~\cite{Daum1997}, the actual image width ($w$) and length ($l$) 
are compared to the expectation value and variance obtained from simulation as a function of the 
image charge $q$ and reconstructed impact distance $\rho$, expressed by two normalized parameters
{\it Scaled Width}~(SW) and {\it Scaled Length}~(SL).

More elaborate analysis techniques were pioneered by the work of the CAT collaboration~\cite{LeBohec1998} on a ``{\it model 
analysis technique}'', where shower images are compared to a  realistic pre-calculated model. Precise description
of the longitudinal development of the shower allowed, for the first time, a bi-dimensional reconstruction even with a single
dish.

\section{The breakthrough - Third generation instruments (CANGAROO-III, H.E.S.S., MAGIC, VERITAS, HAWC)}
\label{sec:thirdgeneration}

The third generation of ground-based telescopes started with the building of 
two new major European collaborations: H.E.S.S. and MAGIC.
The researchers from the CAT, CELESTE and HEGRA experiments 
at the Ecole Polytechnique, CNRS/IN2P3 in Paris and MPI for nuclear physics in Heidelberg
formed the core of the collaboration named
High Energy Stereoscopic System (H.E.S.S.) to build an array of four $12\U{m}$ diameter
IACTs in Namibia. Originally 16 telescopes were planned, and in the first phase 4 IACTs were built.
The first of the four telescopes of Phase I of the H.E.S.S. project went into
operation in Summer 2002; all four were operational in December 2003, and were
officially inaugurated on September 28, 2004. H.E.S.S. started operation in
2002 and is as of now the most successful IACT installation.

The other part of the HEGRA experiment led by MPI for physics in Munich
was joined by several Spanish groups as well as by Italian INFN researchers, to build the core of
Major Atmospheric Gamma-ray Imaging Cherenkov (MAGIC) collaboration to construct a single $17\U{m}$ diameter
IACT in La Palma, the same site as was previously used by HEGRA.  
The first MAGIC telescope was inaugurated in 2003 and the science program started in autumn 2004.

During the same time the two European collaborations were formed, the U.S.-dominated Whipple collaboration proposed to build an 
array of seven $12\U{m}$ diameter IACTs in Arizona, U.S, naming the new project VERITAS.
Four telescopes were finally built at the Whipple Observatory in southern Arizona
and not, as planned, at the Kitt Peak National Observatory. VERITAS went in full operation in 2008.

Around the same time, early 2000's, the Australian-Japanese collaboration CANGAROO upgraded their telescopes for the phase III by
constructing four $10\U{m}$  diameter parabolic shape telescopes in Australia which operated until 2011.


The other three major collaborations underwent a series of upgrades in the recent years.

The MAGIC collaboration constructed a second telescope, almost clone of the first one, 
in 2008; both telescopes have been operating in stereoscopic mode since 2009.
In 2011-2012, the MAGIC telescopes were further upgraded to have finer-grained cameras, larger trigger area and a better readout system
for improved sensitivity.

The H.E.S.S. collaboration constructed a 
much larger fifth telescope - the $28\U{m}$ diameter H.E.S.S. II - placed in the middle of the H.E.S.S. Phase~1 array.
The H.E.S.S.~II telescope has been operational since July 2012, extending the energy coverage towards lower energies and further improving sensitivity.

The VERITAS telescopes were also upgraded: one telescope was moved to a new location for optimizing the sensitivity of the array
in 2009. Further, photomultipliers of all four cameras were changed for more sensitive ones in 2012, thus lowering
the energy threshold of the experiment and increasing its sensitivity.

Lowering the energy of the threshold of the instruments, as aimed by the H.E.S.S.~II, MAGIC and VERITAS telescopes, not only improves the 
overlap with HE space instruments such as Fermi-LAT, but also open new possibilities. High energy emission of pulsars or gamma-ray bursts~\cite{CRAS_GRB}
can for the first time be investigated with ground-based instruments.

\begin{figure}[ht!]
\begin{center}
\includegraphics[width=0.98\textwidth]{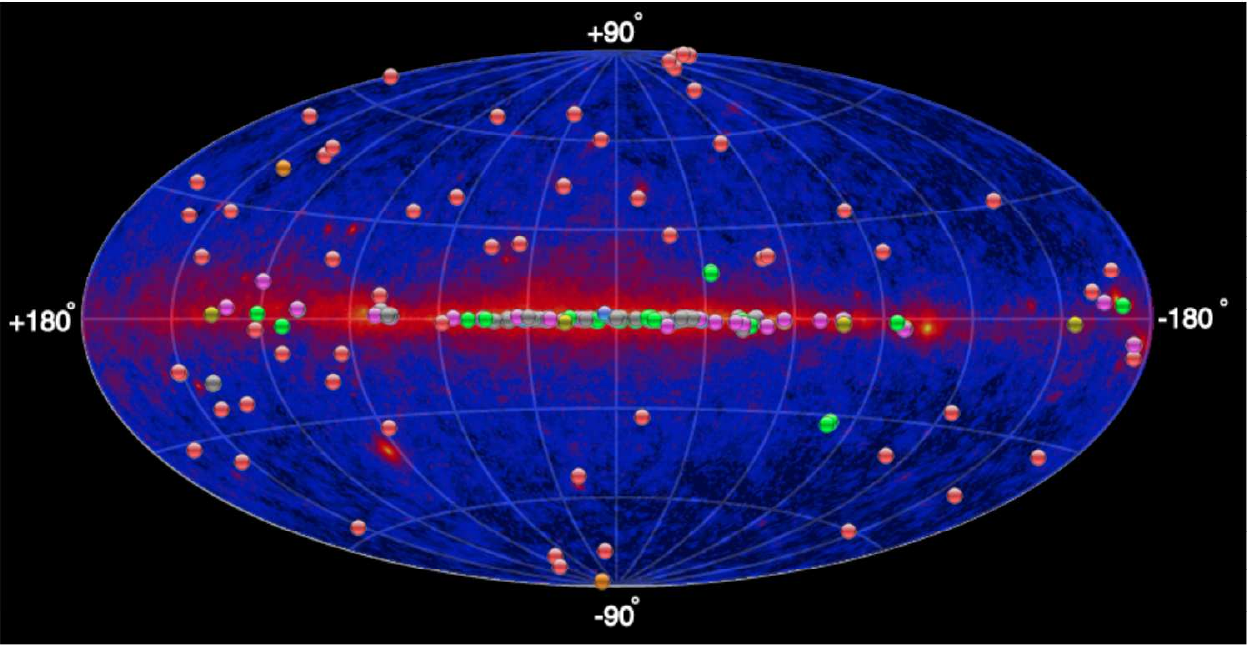} 
\end{center}
\caption{\label{fig:TeVCat}The VHE sky 2015. From {\textit {http://tevcat.uchicago.edu/ .}}} 
\end{figure}

The parameters of the third generation instruments are summarized in Tab.~\ref{Table:IACTs} and the VHE sky map of 2015, resulting mainly from their discoveries,
is shown in Fig.~\ref{fig:TeVCat}.

\begin{table}
\begin{center}
\begin{tabular}{|p{3.200cm}|p{2.4cm}|p{2.4cm}|p{2.4cm}|p{2.4cm}|p{2.4cm}|}
\hline
~
 &
\centering \bfseries CANGAROO III &
\centering \bfseries HESS &
\centering \bfseries MAGIC &
\centering \bfseries VERITAS &
\centering \bfseries HESS-II\cr
\hline
 Number of telescopes &
\centering 4 &
\centering 4 &
\centering 1$\rightarrow $2 &
\centering 2$\rightarrow $ 4 &
\centering 4 (HESS I)+1\cr
\hline
 Dish Diameter (m) &
\centering  10 &
\centering  12 &
\centering  17 &
\centering  12 &
\centering  28\cr
\hline
 Site &
\centering  Australia &
\centering  Namibia &
\centering  Canaries &
\centering  Arizona (US) &
\centering  Namibia\cr
\hline
 Altitude (m a.s.l.) &
\centering  160 &
\centering  1800 &
\centering  2200 &
\centering  1250 &
\centering  1800 \cr
\hline
 Pixels per camera &
\centering  427  (552) &
\centering  960 &
\centering  396+180 $\rightarrow $ 1039 &
\centering  499 &
\centering  2048\cr
\hline
 Pixel Field of View ($^\circ$)&
\centering  0.17 (0.115) &
\centering  0.16 &
\centering  0.1-0.2 $\rightarrow $ 0.1 &
\centering  0.15 &
\centering  0.1\cr
\hline
 Trigger Field of View ($^\circ$)&
\centering  4 (3) &
\centering  5 &
\centering  2.0 $\rightarrow $ 2.6 &
\centering  3.5 &
\centering  3.5 \cr
\hline
 Camera Field of View ($^\circ$) &
\centering  4 (3) &
\centering  5 &
\centering  3.5 &
\centering  3.5 &
\centering  3\cr
\hline
 Readout speed &
 ADC, $100\U{ns}$ integration  &
 $1\U{GHz}$ ARS analog memory, \newline $16\U{ns}$ integration &
 $300\U{MH}$z FADC $\rightarrow $ $2\U{GHz}$ DRS4 &
 $500\U{MHz}$ FADC &
 $1\U{GHz}$ SAM analog memory, \newline $16\U{ns}$ integration \cr
\hline
\end{tabular}
\end{center}
\caption{\label{Table:IACTs}Parameters of third generation instruments. The first CANGAROO-III telescope was slightly different from the other 3, its parameters are in parenthesis.
Upgrades are indicated by an arrow.}
\end{table}

\subsection{IACTs - overall concept}


The concept followed by the major IACT collaborations includes the following four main concepts (developed in chronological order):
\begin{itemize}
 \item Large mirrors in order to collect as much light from low-energy air showers as possible. The energy threshold decreases 
linearly with the mirror area, if the night sky background (NSB) contribution per pixel is kept at a level below 2 photo-electrons per event. 
For cost reasons, the mirror surface of all IACTs is segmented in individual, relatively small $0.5\U{m}$ -- $1\U{m}$ facets.
Two optical arrangements of facets are used, depending on the size of the dish: relatively small ($\sim 10\U{m}$ diameter)
telescopes follow the Davies-Cotton optical design\footnote{The Davies-Cotton design consists in placing
panels of focal length $2 \times f$ on a sphere of focal length $f$, in such a way that, even for inclined rays, some facets are on-axis, thus
reducing the coma aberrations.}, which keeps optical aberrations
at a very low level, even for rather large offsets from the camera center, at the expense of a small anisochronism of $\sim 4\U{ns}$. 
For larger telescopes, parabolic mirror shapes are mandatory to keep the arrival times of the shower photons
isochronous at the focal plane.
 \item Stereoscopic observations: multiple images of the same air shower provide superior gamma/hadron separation,
as well as energy and angular resolutions compared to single telescope images.
 \item Fine-grained cameras with a pixel field of view of the order of $0.1^\circ$ (for a total field of view
of a few degrees) to improve the definition of the shower image, thus leading to an efficient gamma/hadron separation, 
as well as better energy and angular resolutions.
 \item Fast integrating electronics (with a sampling rate larger than $500\times 10^6$ samples per second) that, together 
with the small pixel field of view, keeps the NSB contribution low, which is important for achieving a low energy threshold at the trigger level.
\end{itemize}

VERITAS and H.E.S.S.-I phase telescopes follow the Davies-Cotton optical design, whereas MAGIC and H.E.S.S.-II utilize parabolic mirror shapes.
The CANGAROO-III telescopes also used a parabolic shape reflector. 

The individual segments are spherical mirrors with $0.5\U{m}$--$1\U{m}$ diameter (shape is round, square or hexagonal, depending on individual design),
from several technologies (front aluminized glass, diamond-milled aluminum, etc.
The trigger gate can be kept short for isochronous shower photons, order of several $\U{ns}$, which allows 
the contribution of NSB to be suppressed. In the same time, the time for charge integration in a channel
can be kept short, too, around $3\U{ns}$. This motivates using very fast electronics, short trigger gates 
and short charge integration times, which help significantly in lowering the instrument threshold.
Davies-Cotton telescopes with reasonable focal lengths does not have this possibility due to the spread 
in arrival times of the shower photons because of the spherical mirror
and the use of very fast readout and trigger electronics becomes less appropriate. 
However, in case of a parabolic mirror, the shower images suffer from increased coma aberration,
which degrades the energy resolution and has also negative effects on the trigger because the photons coming from the same sky direction could end up
in different camera pixels, which dilutes the photon density in a pixel.

The trigger systems usually follow usually a three-level concept:
\begin{itemize}
\item A minimal response is required at the single pixel level, usually by means of a simple discriminator (or a constant fraction discriminator).
\item The second level trigger is built as a pattern trigger: either groups of neighbouring pixels
or a given number of pixels within a pre-defined area must fulfill the first level trigger condition.
\item The third level is an array trigger which requires a coincidence between at least two IACTs
within a given time window. 
\end{itemize}

An alternative to the first and second level digital triggers as described above is the so-called sum-trigger.
The sum-trigger adds copies of the PMT signals in an analog way in a certain camera area (patch) and the trigger decision 
for the camera is based upon the strength of the added signal in a patch.
The sum-trigger is more efficient for low energy events. However, it requires additional electronics
to first clip the analog signals from individual pixels in order to remove large after-pulsing signals from PMTs to mimic large 
summed up signals and, secondly, to equalize the timing of the signals for the analog sum at a precision of $1\U{ns}$ for an effective signal pile-up.
A sum-trigger system was installed in the two MAGIC telescopes as an alternative trigger solution after a prototype sum-trigger
proved successful by detecting the Crab pulsar at energies above $25\U{GeV}$ with a single MAGIC telescope.


\subsection{Data mining}

Following the effort carried out in the 1990-2000's, the data analysis methods have been largely improved in the last decade.
Not only  the detectors have been studied in great detail and Monte Carlo simulations improved but also
the analysis methods themselves became quite sophisticated.

The model analysis, initiated by CAT  has been further developed with
a more precise model of showers, which depends on the altitude of the first interaction, the introduction
of a complete modeling of the background using a log-likelihood approach and the extension to stereoscopy~\cite{denaurois2009}.

A completely new analysis technique, ``{\it 3D Model analysis}''~\cite{Lemoine-Model3D,naumann2009} was developed,
based on a 3  dimensional elliptical modeling of the Cherenkov  photo-sphere on the sky, 
that is then adjusted on the observed images simultaneously on all telescopes, thus taking into account intrinsically
the stereoscopic nature of the observation.

In the 1990's simple cuts on basic shower parameters were applied to 
achieve an efficient gamma/hadron separation and simple look-up-tables created to obtain 
best guesses for particle properties.
In the last 5 years the very large increase in computing power led to the advent of massive simulations, which opened
the possibility the reproduce and understand the behaviour of the instruments in much greater details.
Since then, the Random Forest, neural networks and boosted decision trees~\cite{denaurois2006}
are much more popular in use since they proved to improve the separation as well as the angular and energy resolution.
The image cleaning procedures have been steadily improved, too, to make use
of the faint light recorded at the edges of the shower image and to extract it from the NSB. 


The advent of very fast readout systems of the cameras,  
with one or more measurements of the Cherenkov signals per ns in every channel
allowed to record
not only the total intensity in a pixel, but also its time evolution. This opened a completely
new field in analysis technique.
The timing information can be used in single telescope mode to break the degeneracy between distant, high-energy showers
and nearby, lower-energy showers. This was pioneered by the HEGRA experiment~\cite{HESS1999} which first measured
a time gradient along the major axis in the Cherenkov images. The MAGIC telescope~\cite{magic-timing}
more recently demonstrated that, in single telescope mode, the use of timing lead to an improvement 
of the background rejection by a factor of $\sim 2$ due to the combination of two different effects:
\begin{itemize}
\item By integrating the signal only around the time of its maximum, the effect of the night sky background
can be minimized.
\item The time gradient across the field of view can be used as an additional discriminating parameter.
\end{itemize}
Similar studies performed by the VERITAS collaboration in stereoscopic mode~\cite{Holder2005}
did not, however, lead to any improvement beyond the reduction of the night sky background.

The variety of the methods both boosts the sensitivity of the experiments and increases the robustness
of the spectral and morphological reconstruction of gamma-ray sources.

\subsection{Main systematics effects}

The main systematic effects are due to uncertainties in the absolute energy scale 
and in the knowledge of the atmospheric conditions.
The former is related to the fact that there is no calibration beam of VHE gamma rays.
The latter is connected to changes in the temperature, pressure and humidity profiles in the atmosphere,
as well as amount of aerosols or thin clouds: they affect the density of Cherenkov light arriving at the observatory level
and, therefore, influence the energy reconstruction and the trigger efficiency (effective collection area). 
The current best estimates on the energy scale are 
at a precision of 10-15\% and the absolute flux level is uncertain to 10-15\%.

The Earth magnetic field tends to separate apart positively and negatively charged particles in the shower, leading to a broadening
of the shower in the direction orthogonal to the magnetic field. This introduces an asymmetry in the response of the system, with degraded
performances in regions of the sky perpendicular to the direction of the observation. This effect, particularly important at low energy,
is however properly taken into account by Monte Carlo simulations and are therefore not a major part of the systematic uncertainty.

\subsection{New generation particle samplers}

\begin{figure}[ht!]
\begin{center}
\includegraphics[width=\textwidth]{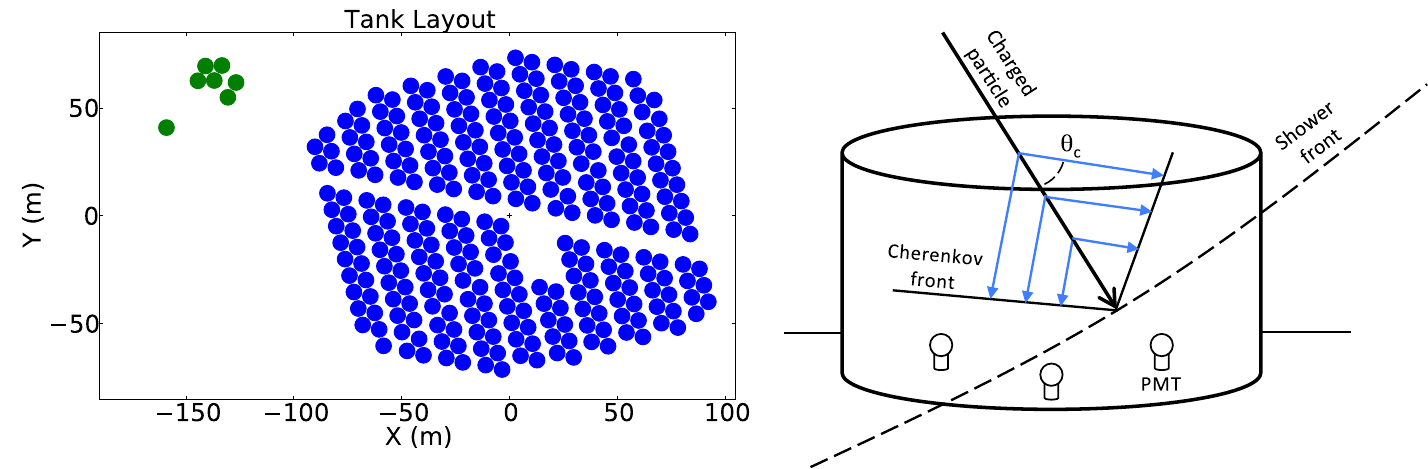}
\caption{\label{fig:HAWC}HAWC layout and operation principle. {\bf Left:} HAWC tank layout. {\bf Right:} Sketch of Water Cherenkov Detection Principle. From~\cite{2012APh....35..641A}.}
\end{center}
\end{figure}

The successor of Milagro, HAWC (High Altitude Water Cherenkov)~\cite{2012APh....35..641A,2013APh....50...26A} relies
on a slightly different concept to its predecessor: instead of a single, large pool, it consists of an array of
300 large water tanks ($7.3\U{m}$ diameter and $4.5\U{m}$ height) instrumented with 4 photomultipliers each (Fig.~\ref{fig:HAWC}). 
Beside simplifying a lot the filtration and
the filling of the system (which can be done step by step), this different geometry provides a much better
hadronic rejection (due to optical isolation between the modules) and a more precise reconstruction.
HAWC, recently inaugurated, is operated in the energy band 
between $100\U{GeV}$ and $100\U{TeV}$ with a sensitivity 15 times better than that of Milagro. This is 
the largest dense particle sampler so far.

\section{Towards large scale, worldwide observatories}
\label{sec:largescale}

\subsection{Challenges}

While the VHE gamma-ray instruments have matured they faced new challenges.
The first is the amount of data that is being taken and needs to be processed and archived.
While the duty cycle of HESS, MAGIC or VERITAS is about 1200\,h per year (which does not seem to be large), 
they record some 200-400\,TB per instrument per year. For the life time of an experiment
(about 10 years is foreseen) it becomes several Petabytes of data.
This made the use of distributed computing and large data centers an essential point in cost, but also 
in successful operation of the instruments.

The second challenge is the automation of the telescope operation.
The observer teams are still required to be on-site to take shifts of data taking
but the instruments are so complex and the operation duty cycle is so demanding
that it became impossible to educate every observer on shift with many technical details
of the instruments. To ensure stable operation and reduce human errors, the automation
of the operation is implemented and the role of the observers is mainly to react in case
automatic procedures fail.

The third challenge is the automation of the data processing. The amount of data
mentioned earlier is so large that it cannot be processed manually.
Automatic pipelines for data quality, data calibration,
low level image parameter calculation as well as for high level products
such as sky maps and energy spectra are implemented. Team of experts are 
overlooking the pipelines to react on possible exceptions and problems
but this must be well organized to reduce the manpower needed.

\subsection{Building a new community}


There is an increasing demand on close cooperation between different 
VHE collaborations on several scientific topics, particularly with the HE community (space-borne detectors).
 Shared efforts for monitoring variable
sources and maximizing time coverage of flaring known sources are some of the examples
of good cooperation. In the extragalactic sky, joint campaigns on the radio galaxy M\,87
discovering a day-scale VHE gamma-ray variability is the best example.
In the Galaxy, joint campaigns on gamma-ray binary LS\,I+61\,303 
trying to unveil the reason for a super-orbital modulation of the emission is another good example.

Though positive examples exist, large collaborations and their rather strict data access and publication rights 
make joint publication a complicated and rather time-consuming process. Experts that are non-collaboration members
do not have access to IACT low level products and are usually discouraged by the size of the collaborations
and the complexity of the data processing. Therefore, a call for a different 
organization of IACTs, such as is common at optical observatories, with open calls for observation proposals and data rights
either partially or completely public, gains a large popularity.
Operating IACTs as open observatories, implying public data analysis tools and data archives
accessible to general astronomers, would necessarily increase the scientific output of the observations
through a participation of astronomers and astrophysicists that are experts on the research field
but non-experts of the IACT data analysis. 

Publication of scientific results becomes another challenge.
At the beginning of the successful era of the VHE gamma-ray community  
every gamma-ray source detection was a great discovery leading to a publication.
Nowadays, and probably even more after the advent of CTA, a well organized 
multi-wavelength campaign generally needs to be conducted and detailed studies of the source 
behavior performed to learn something significantly new. 
While the discovery potential of the instruments is still high
(e.g.\ only about 5\% of the extragalactic sky were observed with IACTs so far)
the effort to produce a scientific publication of the high standard increases.
The publication policy of major IACTs typically requires that the entire collaboration (100-150 people) sign
every scientific publication, in order to acknowledge the high effort of individuals in instrument construction and 
operation. With the increase of collaboration sizes, this model might lead to unacceptable delays and may need
to be revised.


\subsection{Cherenkov Telescope Array}


What was not possible in 1992~\cite{Fleury1992a}, became reality around 2006. 
A community-driven effort resulted in unifying major European, American, and Japanese
groups from H.E.S.S.\, MAGIC and VERITAS to design a large scale Cherenkov telescope observatory.
The maturity of the current big experiments and the need for larger financial resources 
to significantly improve the sensitivity of the instruments led to a consolidation
of the efforts for the next generation instrument. 

The Cherenkov Telescope Array (CTA) was born as an idea for two observatories, one in each
hemisphere, and to cover energies from 10s of GeV to 100s of TeV with unprecedented sensitivity.
The project was funded for the design phase of 5 years in 2008.

At the moment, CTA brings together a community of more than 1500 scientists from all over the world,
with groups in Europe, Asia, North and South-America, Africa, and Australia.
The collaboration is presently nearly ready to start the construction of more than 150 telescopes.

The optimization process of the array layout (in terms of cost and sensitivity) 
favors usage of three different telescope types on the same site: few large size telescopes (LSTs), some 15-25 mid size telescopes (MSTs),
and many (order of 70) small size telescopes (SSTs).
For low gamma-ray energies, the major task is to collect every single Cherenkov photon from faint air-showers.
This motivates usage of LSTs: four $23\U{m}$ diameter parabolic dish telescopes with a field of view of $4.5^\circ$.
The physics of transient phenomena, such as very-high-energy counterparts of Gamma Ray Bursts, requires
fast slewing telescopes which limits the camera weight and therefore prevents the usage of a larger field of view.
LSTs, which essentially follow the MAGIC design with a light weight structure, are most sensitive between $20\U{GeV}$ and few TeV.
The MSTs are an optimization result of HESS-1 (camera and trigger) and VERITAS (mount) telescopes and target
the golden energy regime of IACTs around $1\U{TeV}$. 
With $12\U{m}$ diameter Davies-Cotton reflector, fine-grained cameras and fields of view larger than $7^\circ$,
MSTs will have an improved survey capability. 
The number of MSTs is optimized to obtain multiple ($>4$) images from the 
same shower in the TeV range, which further improves the gamma-ray sensitivity. MSTs are most sensitive between $200\U{GeV}$ and $50\U{TeV}$.
The SSTs are targeting gamma rays with energies above $20\U{TeV}$, where detection and reconstruction is less a problem
(the showers are rather luminous) but the expected fluxes are low. This motivates the construction of many SSTs
in order to cover large area of up to $3\UU{km}{2}$. Both SSTs and MSTs have dual mirror version prototypes,
dubbed SST-dual mirror and Schwarzschild-Couder telescope respectively,
which reduce camera sizes and encourage usage of small size advanced photo-sensors such as Silicon photomultipliers. 
The disadvantage of this design is the shadowing created by the secondary mirror,  leading to a higher energy threshold which, however, 
is not critical due to the targeted energies at which  showers contain a large amount of Cherenkov photons.
 A challenge of the dual-mirror
technique is the precision of the mirror adjustment and control (at some critical points down to few tens of micrometers);
the reward being an improved angular resolution and larger field of view.

Operating a ``{\it Hybrid}'' system, consisting of telescopes of different sizes, is
however a complicated task. The observation strategy needs to be defined very precisely depending on the physics case
(which telescope to use on a specific source, \dots), but also also at the reconstruction and analysis stage.
HESS-II is currently the only hybrid system operating in the world. The data stream consists in both stereoscopic events 
(seen by at least two telescopes out of five) and monoscopic events (seen only by the large telescope). 
The analysis techniques are still under development to be able to cope with this much larger
complexity, and this will become a real challenge for CTA. More and more accurate simulations are needed as 
new, more complex, detectors will come on line.

CTA will in general improve the sensitivity by an order of magnitude compared to what can be achieved with the current generation.
Moreover it will extend the accessible energy range from $20\U{GeV}$ (best overlap with gamma-ray satellites) 
to some 300\,TeV (crucial for Pevatron searches). The cost of the project is about 300\,M Euro and the construction phase should
begin in 2016 with the goal to operate the full array by 2021. The sites of the CTA observatories have not yet been decided,
with Chili and Namibia selected as Southern site candidates and Baja California (Mexico) and La Palma, Canary Islands (Spain)
for the Northern site.

\section{Outlook}
\label{sec:discussion}

\begin{figure}[ht!]
\begin{center}
\includegraphics[width=0.94\textwidth]{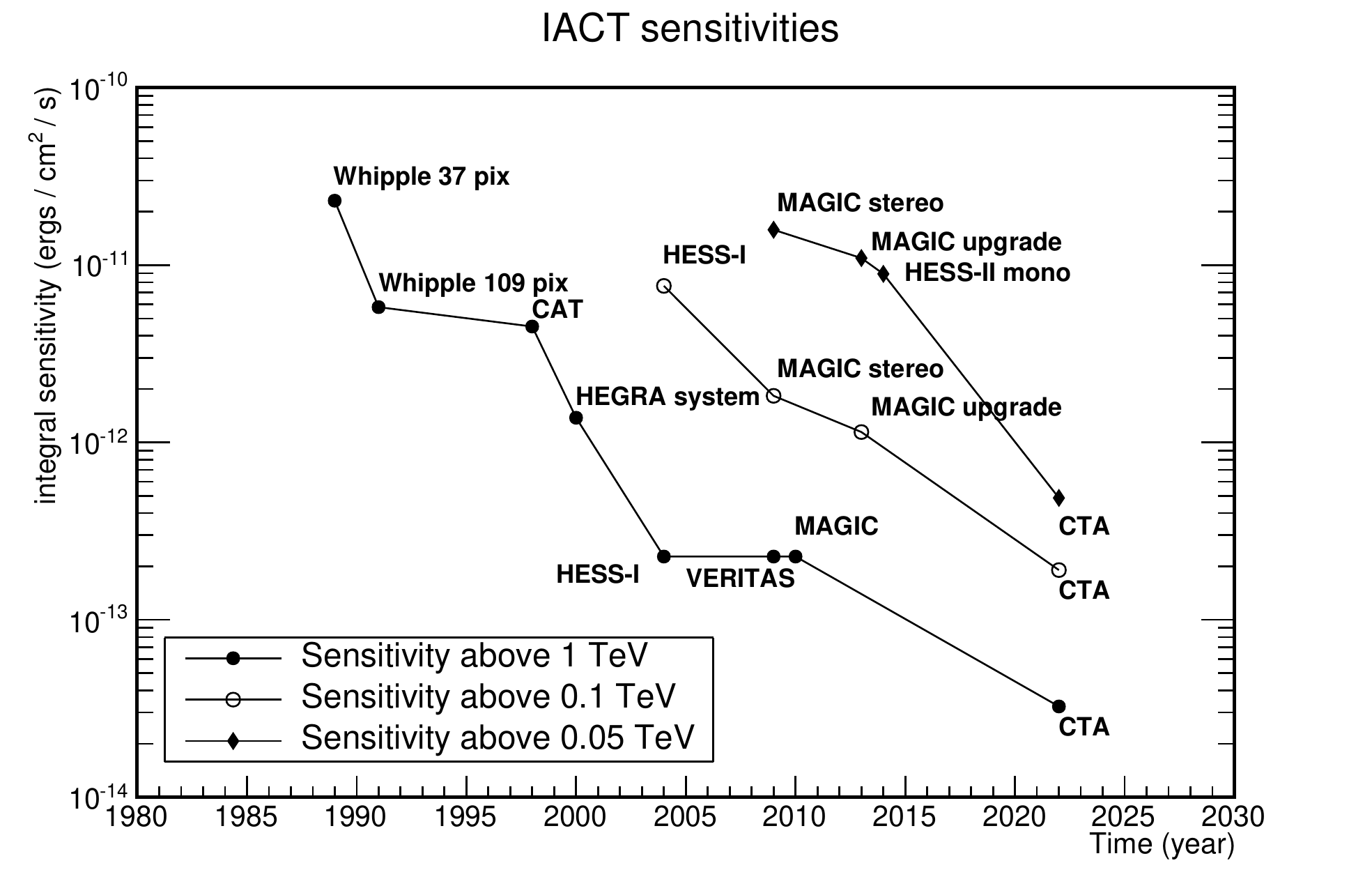} 
\end{center}
\caption{\label{fig:SensVsTime}Evolution of integral sensitivity of the IACTs over time above 
100\,GeV (open circles) and 1\,TeV (filled circles). The expected sensitivities of CTA are also shown. See text for references.}
\end{figure}

\begin{figure}[ht!]
\begin{center}
\includegraphics[width=0.94\textwidth]{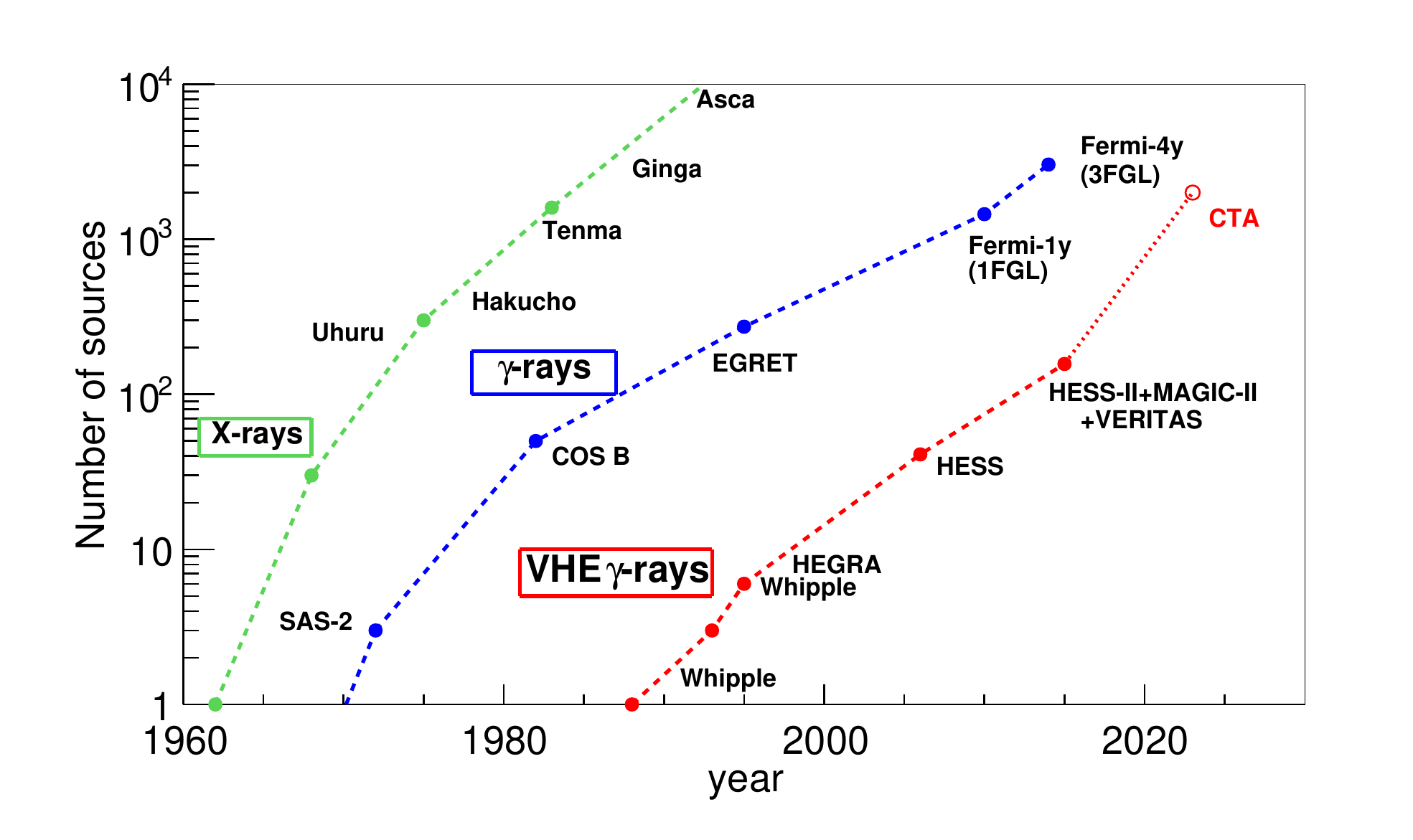} 
\end{center}
\caption{\label{fig:kifuneplot}The so-called Kifune plot showing evolution of number of sources
at different wavelengths as a function of time and different instruments. 
The CTA point is a prediction after 2 years data taking with two full arrays.}
\end{figure}

The IACT technique proved to be very efficient in detecting gamma-ray induced air showers and distinguishing them from the dominating
background, mainly consisting of hadron-induced air showers. The evolution of the integral gamma-ray 
flux sensitivity for point sources is shown in Figure~\ref{fig:SensVsTime}
as a function of time above three different thresholds: $50\U{GeV}$, $100\U{GeV}$ and $1\U{TeV}$.
The sensitivities are computed as Signal/sqrt(Background) for a 50\,h observation
for consistency with older experiments, even if this formula gives 
slightly overestimated results. The numbers are taken or computed from 
\cite{whipple-crab,Vacanti:Crab:1991a,LeBohec1998,Puehlhofer:HEGRA:Perfomance:2003a,hess-crab,aleksic:2012:magic:stereo:performance,aleksic:2014:magic:stereo:performance,Bernloehr:CTA:DesignStudies:2013a} and are sometimes approximate.
Since the first discovery of a VHE gamma-ray source (Whipple observatory, 1989),
the sensitivity of the instruments at energies above $1\U{TeV}$
improved by more than two orders of magnitude in less than 20 years. 
The sensitivity above $100\U{GeV}$ improved by 1.5 orders of magnitude in 10 years.
The extrapolation to the CTA era at all energies accessible by the IACTs 
clearly shows that the sensitivity potential of this technique is far from being saturated.
In the $50\U{GeV}$ to $1\U{TeV}$ energy range, the expected sensitivities are almost an order of magnitude better than the
ones of the current instruments all together.

By having such an enormous improvement in sensitivity in the IACT technique over the last 25 years
there is no surprise that the number of detected sources at VHE
should explode, too.
In fact, the situation of the VHE gamma-ray astronomy is similar to the one of X-ray
or lower energy gamma rays: there is almost an exponential rise of number of sources
after a new window is opened in the electromagnetic spectrum, see Fig.~\ref{fig:kifuneplot}.
The so-called ``{\it Kifune-plot}'', 
named after T. Kifune, who first showed a similar plot at the 1995 ICRC conference in Rome,
shows that the number of detected sources does not saturate in the first 20-30 years
and the CTA simulations follow this development. What this plot also shows
is that for the time being the number of sources is not limited by their 
scarcity but by the sensitivity
of the instruments. 
For many researchers, the recent success of VHE astronomy comes as a surprise, as in the 1980's only a handful
of them  believed there was any detectable gamma-ray source above $100\U{GeV}$.




\bibliographystyle{elsarticle-num} 
\addcontentsline{toc}{part}{Bibliography}
\bibliography{iact-techniques,hess,bibtex_db,whipple,veritas,magic,sampling,algorithms,shower,milagro,hawc,cras} 



\end{document}